\documentclass[twocolumn,superscriptaddress,showpacs,nofootinbib,preprintnumbers,secnumarabic,amssymb, nobibnotes, aps, prd]{revtex4-2}
\usepackage[utf8]{inputenc}
\usepackage{graphicx}
\usepackage{latexsym,amsmath,amssymb,amsthm,lmodern,float,url}
\usepackage{natbib}
\usepackage{color}
\usepackage{microtype}
\usepackage{import}
\usepackage{bbold}
\usepackage[plain]{fancyref}
\usepackage{varioref}
\usepackage{slashed}
\usepackage{multirow}
\usepackage{tikz}
\usepackage{scrextend}
\usepackage{braket}
\usetikzlibrary{shapes}
\usetikzlibrary{positioning}
\usepackage[normalem]{ulem}

\newcommand{\fig}[1]{Fig.~\ref{fig:#1}}
\newcommand{\tab}[1]{Tab.~\ref{tab:#1}}
\newcommand{\eq}[1]{Eq.~(\ref{eq:#1})}

\usepackage[colorlinks=true,backref=false, linktocpage=true,
citecolor=blue,urlcolor=blue,linkcolor=blue,pdfpagemode=UseOutlines]{hyperref}
\hypersetup{%
  bookmarksnumbered=true,
  pdftitle = {},
  pdfsubject = {},
  pdfauthor = {},
  pdfkeywords = {}
}

\DeclareMathOperator{\tr}{Tr}

\begin{document}
\preprint{USTC-ICTS/PCFT-24-06, FERMILAB-PUB-23-570-T}
\title{
Quantum error thresholds for gauge-redundant digitizations of lattice field theories} % Force line breaks with \\
\author{Marcela Carena}
\email{carena@fnal.gov}
\affiliation{Fermi National Accelerator Laboratory, Batavia,  Illinois, 60510, USA}
\affiliation{Enrico Fermi Institute, University of Chicago, Chicago, Illinois, 60637, USA}
\affiliation{Kavli Institute for Cosmological Physics, University of Chicago, Chicago, Illinois, 60637, USA}
\affiliation{Department of Physics, University of Chicago, Chicago, Illinois, 60637, USA}
\author{Henry Lamm}
\email{hlamm@fnal.gov}
\affiliation{Fermi National Accelerator Laboratory, Batavia,  Illinois, 60510, USA}
\author{Ying-Ying Li}
\email{yingyingli@ustc.edu.cn}
\affiliation{Peng Huanwu Center for Fundamental Theory, Hefei, Anhui 230026, China}
\affiliation{Interdisciplinary Center for Theoretical Study,
University of Science and Technology of China, Hefei, Anhui 230026, China}
\author{Wanqiang Liu}
\email{wanqiangl@uchicago.edu}
\affiliation{Department of Physics, University of Chicago, Chicago, Illinois, 60637, USA}

\date{\today}

\begin{abstract}
In the quantum simulation of lattice gauge theories, gauge symmetry can be either fixed or encoded as a redundancy of the Hilbert space. While gauge-fixing reduces the number of qubits, keeping the gauge redundancy can provide code space to mitigate and correct quantum errors by checking and restoring Gauss's law. In this work, we consider the correctable errors for generic finite gauge groups and design the quantum circuits to detect and correct them. We calculate the error thresholds below which 
the gauge-redundant digitization with Gauss's law error correction has better fidelity than the gauge-fixed digitization. Our results provide guidance for fault-tolerant quantum simulations of lattice gauge theories.
\end{abstract}

\maketitle
\section{Introduction}
Gauge symmetries in quantum field theories give rise to extremely rich phenomena. Most prominently, $SU(3) \times SU(2) \times U(1)$ gauge symmetry describes the interactions of the Standard Model. Making ab initio predictions for comparison to experiment requires large computational resources. In particular, Monte Carlo methods in lattice gauge theory (LGT) have been fruitful in the past couple of decades, thanks to the advancement of supercomputers and algorithms. However, problems involving dynamics such as out-of-equilibrium evolution in the early universe~\cite{Polkovnikov_2008, Baum:2020vfl, PhysRevLett.108.080402,Czajka:2021yll}, transport coefficients of the quark-gluon plasma~\cite{Cohen:2021imf} and 
parton physics in hadron collisions \cite{Lamm:2019uyc,Kreshchuk:2020dla,Echevarria:2020wct,Bauer:2021gup,Li:2021kcs,Zache:2023cfj} present sign problems, as the Boltzmann weight becomes complex-valued. 
Future, large-scale quantum computers can avoid this obstacle by performing real-time simulations in the Hamiltonian formalism~\cite{Feynman:1981tf,Jordan:2017lea,Banuls:2019bmf,Bauer:2022hpo}.

In order to use quantum computers for simulations, the infinite-dimensional Hilbert space of the gauge theory must be addressed. To allow a mapping to a finite quantum memory, many digitization proposals to truncate this space have been studied (see Sec VI.b of~\cite{Bauer:2022hpo}). All truncations break the continuous symmetries to some degree and produce theories with smaller symmetries. Understanding the theoretical errors introduced by this is an area of active research~\cite{Shaw:2020udc,Tong:2021rfv,Ji:2020kjk,PhysRevD.106.114504, Bauer:2023jvw,Ciavarella:2023mfc,Hanada:2022pps}. Broadly speaking, methods to encode these regularized theories in quantum computers fall into two classes. The first class digitizes all the states connected through gauge transformations as 
 \textit{redundancy} and uses Gauss's law to project the gauge-invariant subspace, where the physical theory should be simulated. This can be done in group element basis \cite{Zohar:2014qma,Zohar:2016iic,Bender:2018rdp,Gustafson:2022xdt,Hackett:2018cel,Ji:2020kjk,Hartung:2022hoz,Jakobs:2023lpp,Charles:2023zbl,Gustafson:2023kvd}, the group representation basis \cite{Zohar:2012xf,Zohar:2012ay,Zohar:2014qma,Zohar:2015hwa,Singh:2019uwd,Singh:2019jog,Buser:2020uzs}, the mixed basis \cite{Haase:2020kaj}, as a fuzzy gauge theory \cite{Alexandru:2022son,Alexandru:2023qzd}, as a quantum link model \cite{Wiese:2014rla,Luo:2019vmi,Brower:2020huh,Mathis:2020fuo}, and more. The second class digitizes only gauge-invariant states. We refer to this as \textit{fixing the gauge} in the sense that it eliminates the redundant degrees of freedom. One can digitize the independent Wilson loops in the lattice, which can be identified as the plaquettes in $(2+1)$ dimensions \cite{Zohar:2013zla,Kaplan18_GaussLaw, Bender20_compactQED,Yamamoto:2020eqi,Bauer:2021gek,Grabowska:2022uos,Kane:2022ejm,PRXQuantum.2.030334}, and as the states outside the maximal tree in higher dimensions~\cite{Bauer:2023jvw}.  Other digitizations to eliminate redundancy include the Fock basis in the light-front Hamiltonian \cite{Kreshchuk:2020kcz,Kreshchuk:2020dla}, the local multiplet basis \cite{Klco:2019evd,Ciavarella:2021nmj}, and spin networks~\cite{Zache:2023dko,Hayata:2023bgh}. Besides reducing the qubit cost, gauge fixing also simplifies the quantum state preparation, but can complicate the Hamiltonian~\cite{Zohar:2013zla,Zohar:2011cw,Banerjee:2012xg,Tagliacozzo:2012df,Raychowdhury:2018osk,Surace:2019dtp,Davoudi:2020yln,Zache:2023dko,Bauer:2023jvw,Klco:2019evd,Ciavarella:2021nmj,Hayata:2023bgh}.  At present, it is an open question whether gauge fixing is ultimately advantageous for quantum simulation. 

The effect of noisy hardware has been generally neglected in the above discussion. Naively, gauge-redundant digitizations suffer more severely from quantum noise because more qubits introduce more errors. However, not all the errors are equally harmful. Noise which breaks symmetries can change the universality class simulated~\cite{Singh:2019jog,Singh:2019uwd,Bhattacharya:2020gpm,Zhou:2021qpm,Caspar:2022llo,Alexandru:2022son} and thus sufficient symmetry preservation is crucial~\cite{Stryker:2018efp,Raychowdhury:2018osk,Lamm:2020jwv,Mathew:2022nep,VanDamme:2021njp,Halimeh:2022mct}. In the gauge-redundant digitization, quantum errors can be classified as gauge-preserving or gauge-violating operators~\cite{Halimeh:2022mct,PhysRevLett.125.240405,VanDamme:2020rur,Bonati:2021vvs,Bonati:2021hzo,Gustafson:2023swx}. The gauge-violating errors can be mitigated by introducing energy penalties~\cite{Stannigel:2013zka,Halimeh:2020ecg,Halimeh:2020xfd, Halimeh:2021vzf,Tran:2020azk} or random gauge transformations~\cite{Lamm:2020jwv} in the Hamiltonian evolution. They can also be corrected by measuring and restoring Gauss's law ~\cite{Stryker:2018efp,Rajput2023npj}. The gauge-fixed digitization -- while precluded from noise-based gauge violations-- does not have such natural error mitigation or correction methods, and thus relies on generic methods for the residual errors.

 Notably, redundancy and symmetry play a central role in both quantum error correction (QEC) and mitigation (QEM). QEC deliberately designs a redundant full Hilbert space $\mathcal{H}_{\mathrm{full}}$ on the physical qudits and encodes quantum information in a much smaller code subspace $\mathcal{H}_{\mathrm{code}}$ on logical qudits with certain symmetries, thus allowing for correction without disrupting the coherent quantum information in $\mathcal{H}_{\mathrm{code}}$ \cite{Terhal:2015qec,roffe2019quantum}. As an active field of research, estimates for the overhead -- the physical to logical qubit ratio -- vary from $\mathcal{O}(10)$ to $\mathcal{O}(10^5)$~\cite{ionq_2020,ibm_2023, google_2020}. QEM uses the existing symmetries without introducing redundancy to mitigate errors~\cite{cai2022quantum}. Indeed, it is this structural similarity between QEC, QEM and LGT that inspires the above mentioned works~\cite{PhysRevLett.125.240405,Bonati:2021vvs,Bonati:2021hzo,Gustafson:2023swx,Stannigel:2013zka,Tran:2020azk,Lamm:2020jwv,Stryker:2018efp,Rajput2023npj,Mathew:2022nep,VanDamme:2021njp,Halimeh:2022mct} to use gauge redundancy as a resource for error correction and mitigation.  Given the huge variance of the overhead in QEC, and that gauge fixing saves logical qubits only by a factor of approximately $(1-1/d)$ in $d$ spatial dimensions, keeping the gauge redundancy for QEC or QEM may be more resource efficient for achieving a desired accuracy. This idea has only just begun to be explored for field theories~\cite{Klco:2021jxl,Rajput:2021trn,delPino:2022zzx,Gustafson:2023swx} including fermionic systems~\cite{Landahl:2021adh,chen2022errorcorrecting}.

The answer to \textit{when} this is true can be phrased as a \textit{threshold theorem} of QEC \cite{Shor:1996qc, Aharonov2008fault-tolerant,
Knill1998Resilience}, which states that there is a threshold for the error rate of physical qubits, below which more redundancy makes the code more error-proof. 
In this work, we compute the threshold below which gauge redundancy makes the digitization more robust. After reviewing the connection between gauge symmetry and QEC in Sec.~\ref{sec:gauge_sym} and Sec.~\ref{sec:QEC}, we present the circuits to encode and decode via Gauss's law. This paves the way for calculations in Sec.~\ref{sec:fidelity} of the thresholds, below which
the gauge-redundant digitization combined with QEC has a better fidelity than the gauge-fixed one.

\section{Gauge symmetry and fixing} \label{sec:gauge_sym}

We will briefly review gauge symmetry on lattice in both the group element (magnetic) and the group representation (electric) basis, as the former is closer to the path-integral quantization and the latter to Gauss's law in classical fields. These bases are related by the group Fourier transform. In the group element basis, one assigns an element of the group $G$ (link variable) to each link on the lattice, representing the Wilson line~\cite{PhysRevD.11.395}. For continuous gauge groups, the link variables are related to the vector potentials of the continuum theory via:
\begin{equation}
    U_{\mathbf{x},i}=P \exp (-i \int_{\mathbf{x}}^{\mathbf{x+i}}  d \mathbf{l}\cdot \mathbf{A}) \approx e^{-i  a A_i(\mathbf{x})},
\end{equation}
where $P$ denotes path ordering, $a$ is the lattice spacing, and $i$ the spacial direction of the link. The gauge-redundant Hilbert space  $\mathcal{H}_\mathrm{gauge}$ is the tensor product of $N_L$ local spaces, each spanned by the group elements:
\begin{align}
    \mathcal{H}_\mathrm{gauge}=\mathrm{span}(\{\ket{U}, U \in G\} )^{\otimes N_L}
\end{align}
where $N_L$ is the total number of links and $\braket{U|U'}=\delta_{U',U}$.

For a continuous gauge group, $\mathrm{dim}(\mathcal{H}_{\rm gauge})=\infty$, and digitization is required to render it finite. Here we will focus on the discrete subgroup digitization. Some discussion of the continuous theory can be found in~\cite{bao2023superselection} and other digitization schemes remain for future works. For a discrete gauge group $G$, the dimension of the one-link Hilbert space is $|G|$. The dimension of $\mathcal{H}_{\mathrm{gauge}}$ is given by:
\begin{equation}
    \mathrm{dim}\,\mathcal{H}_{\mathrm{gauge}}=|G|^{N_L}.
\end{equation} 

A gauge transformation $\hat T_{g_\mathbf{x}}$ is a unitary operator that transforms all out-going (in-coming) links connected to a site $\mathbf x$ by a left (right) product with $g \in G$ ($g^{-1} \in G$): 
\begin{align}\label{eq:T_g}
    \hat T_{g_\mathbf{x}}=\prod_{i=1}^d[\hat{L}_{g}(\mathbf {x}, i) \hat{R}_{g^{-1}}(\mathbf {x-i}, i) ],
\end{align}
where $\hat{L}_{g},\hat{R}_{g}$ are left and right multiplication operators, 
\begin{align}\label{eq:Left}
    \hat{L}_{g}=\sum_{U \in G} \ket{gU}\bra{U},\, 
     \hat{R}_{g^{-1}}=\sum_{U \in G} \ket{Ug^{-1}}\bra{U}.
\end{align}
Typical lattice gauge Hamiltonians preserves gauge symmetry as they commutes with the gauge transformation operator $\hat T_{g_\mathbf{x}}$. This includes the Kogut-Susskind~\cite{PhysRevD.11.395}, the Symanzik-improved~\cite{Luo:1998dx,Carlsson:2001wp, PhysRevD.64.094503,PhysRevLett.129.051601} and the Similarity-Renormalization-Group-improved ones~\cite{Ciavarella:2023mfc}.

Gauge-invariant states satisfy a lattice version of Gauss's law:
 \begin{equation}\label{eq:Gauss}
   \hat T_{g_\mathbf{x}}\ket{\Psi_{\mathrm{inv}}}=\ket{\Psi_{\mathrm{inv}}}, \forall  \,  g_{\mathbf{x}} \in G.
\end{equation}
In the $\ket{U}$ basis, Gauss's law requires the wave functions on gauge-equivalent configurations to be the same, as taking an inner product of \eq{Gauss} with $\bra{U}$ gives
\begin{align}\label{eq:inv_wf}
    \braket{U|\hat T_{g_\mathbf{x}}|\Psi_{\mathrm{inv}}}=\braket{U|\Psi_{\mathrm{inv}}}
\end{align}

Such states are in the gauge-invariant subspace $\mathcal{H}_{\mathrm{inv}}$, which can be projected from $\mathcal{H}_{\mathrm{gauge}}$ with the operator $\hat{P}_{\mathrm{inv}}\equiv\prod _{\mathbf{x}}\hat{P}_0 (\mathbf{x})$,  where the local Gauss's law projector is
\begin{align}\label{eq:PGauss}
    \hat{P}_0 (\mathbf{x})&= \frac{1}{|G|}\sum_{g_{\mathbf{x}}\in G}\hat T_{g_\mathbf{x}}.
\end{align}

To get a clearer physical picture, it is useful to introduce the electric basis  $\ket{\sigma_{mn}}$, which is the group Fourier transform of the group element basis:
  \begin{equation}\label{eq:Fourier}
    \braket{\sigma_{mn}|U}= \sqrt{d_\sigma /|G|}\Gamma^{(\sigma)}_{mn}(U) ,
  \end{equation}
  where $\Gamma^{(\sigma)}_{mn}(U)$ is the $m,n$ matrix element of the unitary irreducible representation (irrep) $\sigma$ for $U$, $d_\sigma$ is the dimension of $\sigma$. The dimension of the one-link Hilbert space $|G|$ also equals $\sum_\sigma d_\sigma^2$.
  For the Abelian groups, all irreps are 1d and $m,n$ can therefore be suppressed with $\sigma$ being the Abelian electric flux. For non-Abelian groups, $\ket{\sigma_{mn}}$ is a tensor representation  comprised of the left vector representation $\ket{\sigma_m}$ and the right vector representation $\ket{\bar\sigma_n}$:
  \begin{equation}
      \ket{\sigma_{mn}}=\ket{\sigma_m}\otimes\ket{\bar\sigma_{n}},
  \end{equation} 
  where $\bar{\sigma}$ is the complex conjugate dual of representation $\sigma$, i.e. $\Gamma^{(\bar{\sigma})}_{nq}(g)=\Gamma^{(\sigma)}_{qn}(g^{-1})$. This is because gauge transformations act on the links going out from the vertex via 
  \begin{align}\label{eq:L_Ebasis}
    \hat{L}_{g}=\sum_{\sigma,m,q}\Gamma^{(\sigma)}_{mq}(g)\ket{\sigma_{m}}\bra{\sigma_{q}} \otimes \sum_n\ket{\bar\sigma_n}\bra{\bar\sigma_n},
\end{align} 
thus transforming only the left vector. Similarly, the in-coming links transform in the dual representation, and only the right vectors are transformed: 
\begin{align}
\hat{R}_{g^{-1}}=\sum_{\sigma,m}\ket{\sigma_m}\bra{\sigma_m}\otimes\sum_{n,q}\Gamma^{(\bar{\sigma})}_{nq}(g)\ket{\bar\sigma_{n}}\bra{\bar\sigma_{q}}.
\end{align}
Thus, one can define the local charge at vertex $\mathbf{x}$ as the tensor product representation:
\begin{align}\label{eq:charge}
    \ket{Q(\mathbf{x})}= \prod_{i=1}^{d}\otimes \ket{\sigma_m(\mathbf{x},i)} \otimes \ket{\bar{\sigma}_n(\mathbf{x-i},i) }
\end{align} 
$\hat{P}_0 (\mathbf{x})$ projects the tensor product to the trivial representation according to the Clebsch-Gordan coefficients of the group. Physically, one can interpret $Q(\mathbf{x})$ as the net flux, and $\hat{P}_0 (\mathbf{x})$  selects states with neutral net flux. As the result of the constraints, one lacks the freedom to independently assign all the links variables. Naively, applying $\hat{P}_0(\mathbf{x})$ to all the $N_V$ vertices in the lattice implies $N_V$ different constraints. However, the global charge, defined as the tensor product of all the local charges,
\begin{equation}\label{eq:ebasis_global}
    \ket{Q_{\mathrm{gl}}}=\prod_\mathbf{x}\otimes\ket{Q(\mathbf{x})}=\prod_{\mathbf{x},i}\otimes \ket{\sigma_m(\mathbf{x},i)} \otimes \ket{\bar{\sigma}_n(\mathbf{x},i) }
\end{equation}
is not a gauge constraint but rather the physically-conserved charge of the state under the global transformation with $h\in G$:
\begin{align}\label{eq:global}
    \hat{T}_{h}^{\rm{Gl}} \prod _{\mathbf{x},i}\ket{U_{\mathbf{x},i}}= \prod _{\mathbf{x},i}\ket{hU_{\mathbf{x},i}h^{-1}}.
\end{align}

In the case of an Abelian group like $U(1)$ or $\mathbb{Z}_N$, the links cannot carry charge as the charges of the left and right vector always cancel, making the global charge automatically neutral. This can also be easily seen in the group element basis as all states are invariant under \eq{global}. In contrast, non-Abelian gauge fields can carry charge, as indicated by the difference between the left and right vectors, and the non-trivial global transformations. If one allows for all the possible global charges, the number of independent constraints is thus $N_V-1$. This makes $N_L-N_V+1$ links dynamical with the rest $N_V-1$ links dependent on the dynamical ones. Thus,
 \begin{equation}\label{eq:dim_inv}
 \mathrm{dim }\mathcal{H}_{\mathrm{inv}}=|G|^{N_L-N_V+1}. \end{equation}
If one fixes the global transformations as well, thereby selecting the subspace with neutral global charge, the dimension of $\mathcal H_{\mathrm {inv}}$ might be reduced by a factor greater or equal to $|H|/|G|$ where $H$ is the Abelian center of the group, as derived in Appendix \ref{app:orbits}.

\begin{figure}
  \centering
\includegraphics[width=0.47\linewidth]{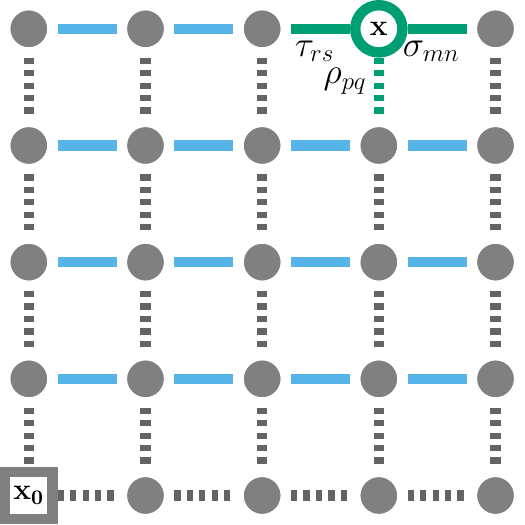}\hspace{0.3cm}
\includegraphics[width=0.47\linewidth]{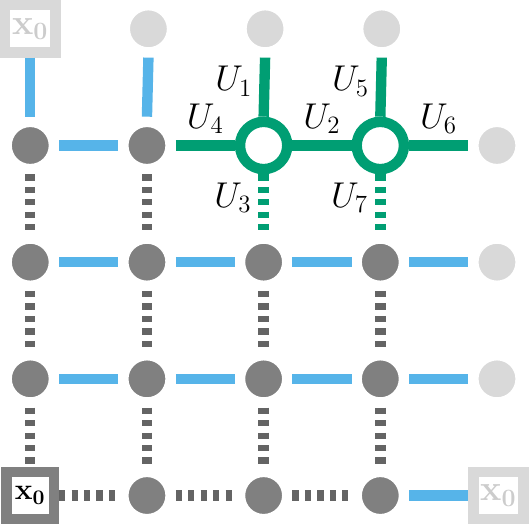}
   \caption{Maximal trees (dashed lattice links) of lattices with (left) OBC and (right) PBC. The site at the lower left corner is chosen as the root of the maximal tree.
   } 
   \label{fig:maxtree}
\end{figure}

 Choosing which degrees of freedom are independent and eliminating others is equivalent to gauge fixing.  On the lattice, the maximal tree method fixes the gauge up to the global transformations \eq{global}, by explicitly solving the local Gauss's law constraints which forces some links to be functions of the remaining ones~(See \cite{Gattringer:2010zz} for a good discussion). For example, consider the links connected to the vertex $\mathbf{x}$ shown in the left of \fig{maxtree}, the state are given by \footnote{We have suppressed the vector indices of representations here for simplicity.}
\begin{equation}
   \ket{\Psi_\mathrm{gauge}} =\sum_{\sigma, \rho, \tau}a(\sigma,\rho,\tau)\ket{\sigma,\rho,\tau}.   
\end{equation} 
Constraining $\ket{Q(\mathbf{x})}$ to be in the trivial representation, the variable $\rho$ in the maximal tree can be determined as $\rho= \sigma \otimes \bar{\tau}$, which produces the gauge-invariant state:
\begin{eqnarray}
\label{eq:instate}
    \ket{\Psi_{\mathrm{inv}}}  
    &=& \sum_{\sigma,  \tau}a(\sigma,\tau)\ket{\sigma,\tau} \otimes \left|\rho= \sigma \otimes \bar{\tau}\right>.
\end{eqnarray}
The same procedure can be repeated till the dependent links form a maximal tree in the lattice, i.e. a set of links which contains one and only one path between any pair of vertices, the size of which is $N_T=N_V-1$.
We will define the Hilbert space spanned by the variables completely determined by Gauss law (in the maximal tree in the above example) as $\mathcal{H}_{\mathrm{red}}$, and the rest as $\mathcal{H}_{\mathrm{fixed}}$. Clearly,
\begin{align}\label{eq:H_inv}
      \mathcal{H}_\mathrm{inv}\cong \mathcal{H}_{\mathrm{fixed}}=\mathrm{span}(\{\ket{U}, U \in G\} )^{\otimes N_L-N_V+1}.
 \end{align}

Similar to choice of gauge fixing in the continuum, for a given lattice there is no unique maximal tree. One popular method~\cite{Ligterink_2000,MULLER198449,Bronzan1985,Bauer:2023jvw} is to pick a site $\mathbf{x}_0$, and the set of links which uniquely connects $\mathbf{x}_0$ via a comb-like path to any other site form a tree of size $N_T=N_V-1$ . This maximal tree is shown in \fig{maxtree} for 2d with open (OBC) and periodic (PBC) boundary conditions. Notably, the charge at $\mathbf{x}_0$ is not fixed by the maximal tree, but left to carry the charge under global transformations in \eq{global} \cite{Bronzan1985,Bauer:2023jvw}.
 
In the \textit{gauge-redundant} quantum simulations digitized with Hilbert space $\mathcal{H}_{\mathrm{gauge}}$ and a quantum computer made of qudits of dimension $N$, the number of qudits required is $\log_N (\mathrm{dim}\,\mathcal{H}_{\mathrm{gauge}})= N_L\log_N |G| $.  \textit{Gauge fixing} eliminates the redundancy by digitizing the subspace $\mathcal{H}_{\mathrm{inv}}$ instead, thus reducing the number of qudits to $\log_N (\mathrm{dim}\,\mathcal{H}_{\mathrm{inv}})$.  For the rest of the work, we take $\log_N (\mathrm{dim}\,\mathcal{H}_{\mathrm{inv}})\approx(N_L-N_T)\log_N |G|$, which is accurate for fixing the maximal tree in both the Abelian and non-Abelian cases, and an upper bound for non-Abelian $G$ if the global symmetry is also fixed. The numbers of sites, links, links in a maximal tree are listed in \tab{DOF} for different boundary conditions.

\begin{table}[h]
\caption{The number of degrees of freedom in a lattice with $L$ links per side in $d$ dimensions. $b=0, 1$ for PBC and OBC respectively.}
\label{tab:DOF}  
\centering
\begin{tabular}{c|c}
    \hline 
    Sites $N_V$ & $(L+b)^d$ \\
     Links $N_L$ & $d L(L+b)^{d-1}$\\
     Maximal Tree Links $N_T$ & $(L+b)^d-1$ \\
     \hline
\end{tabular}
\end{table}

Gauge fixing reduces the number of qubits, but often at the price of complicating the lattice gauge Hamiltonian $\hat{H}_{\mathrm {LGT}}$. Written with the gauge-redundant degrees of freedom, $\hat{H}_{\mathrm {LGT}}$ only consists of local electric and magnetic energy operators. In the gauge-fixed formalism with Hamiltonian $\hat{H}_{\mathrm{fixed}}$, the electric operators of the redundant links in the maximal tree cannot be discarded, but are non-local combinations of operators outside the tree~\cite{Bronzan1985,Bauer:2023jvw}. 
Thus $\hat{H}_{\mathrm{fixed}}$ is generically denser and non-local. It is easier to fix the gauge while keeping the Hamiltonian simple in 1d or 2d.
In 2d, $\log_{|G|}\mathrm{dim}\,\mathcal{H}_{\mathrm{inv}}$ is
\begin{align}\label{eq:Euler_chi}
    N_L-N_T=N_L-N_V+1=N_P-\chi+1,
\end{align}
where $N_P$ is the number of plaquettes, and $\chi$ is the Euler characteristic. This also follows from counting the number of plaquettes with independent magnetic fluxes (Wilson loops), plus the topological degrees of freedom.
For an OBC lattice ($\chi=1$), all $N_P$ plaquettes are independent and there are no topological degrees of freedom.
This plaquette magnetic flux basis keeps the Hamiltonian local and sparse for Abelian gauge groups~\cite{Zohar:2013zla,Kaplan18_GaussLaw, Bender20_compactQED,Yamamoto:2020eqi,Bauer:2021gek,Grabowska:2022uos,Kane:2022ejm,Haase:2020kaj,PRXQuantum.2.030334}.
For a PBC lattice ($\chi=0$), the total magnetic flux is zero, thus there are $(N_P-1)$  plaquettes with independent magnetic fluxes.  Additionally, considering the two independent Wilson loops with winding number one, this yields $\mathrm{dim}\,\mathcal{H}_{\mathrm{inv}}=|G|^{N_P+1}$.

\section{Lattice gauge symmetries as stabilizer codes}\label{sec:QEC}
\begin{figure}
    \centering
\includegraphics[width=\linewidth]{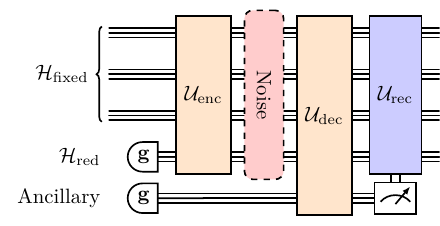}
\caption{Schematic circuit of LGT as QEC. $\textbf g$ represents a color-neutral state.}
    \label{fig:concept}
\end{figure}
In this section, we construct quantum circuits to use gauge redundancy as partial error correction codes for generic finite groups. The method is based on the similarity between stabilizer codes and lattice gauge theories, which we review in \ref{sec:stabilizers}. Without redundancy, any error in $\mathcal{H}_{\mathrm{fixed}}$ causes irrecoverable information loss. In the gauge-redundant formalism, as a result of gauge symmetry  \eq{inv_wf},  all the states equivalent under gauge transformation carry copies of the same wave function. This allows one to detect and correct some gauge-symmetry violating errors, as is analyzed in \ref{sec:correctables}.  
\fig{concept} outlines the circuit to realize such error corrections. The preparation of gauge invariant states is a process to encode the quantum information with the maximal-tree redundancy. To decode, we introduce an ancillary register at each site to compute and measure the net flux. Nonzero fluxes indicates occurrence of quantum errors which can be mitigated via post-selection, with the associated sampling overhead exhibiting exponential growth in the system size \cite{cai2023quantum}. One can also correct these errors by applying the optimal recoveries after deducing the most likely errors. The details of the construction are in \ref{sec:circuit}.
\subsection{Stabilizer codes} \label{sec:stabilizers}
Stabilizer codes perform error correction by using a redundant \textit{full Hilbert space} $\mathcal{H}_{\mathrm{full}}$ of physical qubits (qudits), but encoding quantum information only in states with certain symmetries. The symmetry group that defines the code states is called the stabilizer group $\mathcal{S}$, usually a subgroup of the (generalized) Pauli group on the physical qubits (qudits). A code state satisfies 
\begin{equation}\label{eq:stabilizer}
    \hat{s}\ket{\psi}_{\mathrm{code}}=\ket{\psi}_{\mathrm{code}}, \forall \, \hat{s} \in \mathcal{S}.
\end{equation}
All code states are in the \textit{code space}, $\mathcal{H}_{\mathrm{code}}$. The sum of all $\hat{s}$ is proportional to the projector from $\mathcal{H}_{\mathrm{full}}$ to $\mathcal{H}_{\mathrm{code}}$:
\begin{equation}\label{eq:Pcode}
   \hat{P}_{\mathrm{code}}=\frac{1}{|\mathcal{S}|}\sum_{\hat{s} \in \mathcal{S}} \hat{s}.
\end{equation}

Note that \eq{stabilizer} is similar to \eq{Gauss}, and \eq{Pcode} to \eq{PGauss}. Applying the above concepts to LGT, the group of gauge transformations $\mathcal{G}$ is a  stabilizer group, and $\mathcal{H}_{\mathrm{inv}}$ corresponds to the code space.  The projection operator $\hat{P}_{\rm inv}$ from $\mathcal{H}_{\mathrm{gauge}}$ to $\mathcal{H}_{\mathrm{inv}}$ is the (normalized) sum of all stabilizers.
  \footnote{We will reserve the terms ``physical states (space)" for the gauge-symmetric ones in the LGT which is unfortunately different from the convention of the QEC literature. Terms ``physical qubits (qudits, noise)" still refer to any qubits (qudits, noise) in $\mathcal{H}_{\mathrm{full}}$, consistent with QEC literature.} Measuring whether and how gauge symmetry is violated allows one to detect and correct gauge-violating errors. 
  
In active QEC, one measures a minimum generating set of $\mathcal{S}$, and the results are \textit{syndromes}. To avoid collapsing the quantum information in the measurement of syndromes, ancillae are needed to compute the syndromes coherently:
\begin{align}
    \big[\sum_{X}\psi (X)\ket{X}\big] \otimes \ket{0}_{\mathrm{anc}}
    \to \sum_{X}\psi (X)\ket{X}
    \otimes \ket{s(X)}_{\mathrm{anc}},
\end{align}
where $\ket{X}$ are base vectors of the physical qubits. In the case of Pauli stabilizers with eigenvalue $\pm 1$, measuring the ancillary qubits collapses the state to $\mathcal{H}_{\mathrm{code}}$ when all syndromes return $+1$, or one of the subspaces orthogonal to $\mathcal{H}_{\mathrm{code}}$ if some syndromes return $-1$ indicate the errors. One then applies an optimal recovery operation to the quantum state depending on the syndromes. Clearly, errors that transform one state to another within $\mathcal{H}_{\mathrm{code}}$ are not detectable, and those are \textit{logical errors}.

The \textit{code distance} is the minimum number of single physical qubit (qudit) errors that are part of a logical error. For scalable quantum error correction codes, which imply that the code distance can be arbitrarily large, there exists an \textit{error threshold} for the error rate per physical qudit. If the physical error rate is below such threshold, we can decrease the logical error rate by increasing the code distance, despite the increased total errors due to more physical qudits \cite{Shor:1996qc,Knill1998Resilience,Aharonov2008fault-tolerant}.

\subsection{The set of correctable errors} \label{sec:correctables}
 For a quantum system represented as a density operator $\rho$, the general noisy evolution can be written as a completely positive, trace-preserving map $\mathcal{N}(\rho)$:
\begin{align}
   \mathcal{N}(\rho)=\sum_j \hat N_j \rho \hat N_j^\dagger
\end{align}
where $\hat N_j$ are error operators, usually including the identity operator. 
The necessary and sufficient condition for correctable errors is the Knill-Laflamme (KL) condition~\cite{KnillQECC97}:
\begin{align}\label{eq:KL}
     \hat{P}_{\mathrm{code}} \hat{N}_i^\dagger \hat{N}_j \hat{P}_{\mathrm{code}} =\lambda_{ij}  \hat{P}_{\mathrm{code}}, \quad\lambda_{ij}^\ast  =\lambda_{ji} .
\end{align}

For errors as operators on a $G$-register, a linearly independent complete basis can be constructed as $\{\hat{L}_g \hat{\Gamma}_\sigma\}$,  where $\hat{L}_g $ is the group left multiplication operator defined in \eq{Left}, and $\hat{\Gamma}_\sigma$ is the matrix element operator: \footnote{$\Gamma_\sigma$ is short for $\Gamma_{\sigma,m,n}$, where $\sigma$ is the irrep, and $m,n$ indicated the matrix element in the representation. The number of different $\Gamma_{\sigma, m,n}$ is therefore $\sum_\sigma {d_\sigma}^2=|G|$.}
  \begin{align}
      \hat{\Gamma}_{\sigma,m,n}=\sum_{h\in G} \sqrt{d_\sigma} \Gamma^{(\sigma )}_{mn}(h)^\ast\ket{h}\bra{h}
      \label{eq:GM}
  \end{align}
  One can check that the above set of operators are linearly independent by satisfying:
  \begin{align}
      \tr \big [ (\hat{L}_g \hat{\Gamma}_\sigma)^\dagger(\hat{L}_{g'} \hat{\Gamma}_{\sigma'})]=|G| \delta_{g,g'}\delta_{\sigma,\sigma'}.
  \end{align}
  The basis is complete as there are only $|G|^2$ independent operators in a Hilbert space of dimension $|G|$. 

The gauge-violating effect of $\hat{\Gamma}_\sigma$ is clearer in the electric basis. Using \eq{3Gamma} between Clebsch-Gordan coefficients and the matrix elements of irreps in \cite{Broek78CG}, one can derive:
  \begin{align}
     &\bra{\sigma''_{lr}} \hat{\Gamma}_{\sigma,m,n}\ket{\sigma'_{kq}}
=\notag\\&\sqrt{\frac{d_\sigma d_{\sigma'}}{d_{\sigma''}}}  \sum_{\alpha}\braket{ \sigma''_{l},\alpha |\sigma'_{k},\sigma_m}\braket{\bar\sigma''_{r},\alpha |\bar\sigma'_{q},\bar\sigma_n} ,
  \end{align}
where $\braket{ \sigma''_{l},\alpha |\sigma'_{k},\sigma_m}$ is the Clebsch-Gordan coefficient, and the sum of $\alpha$ is from $1$ to the multiplicity $M(\sigma''; \sigma',\sigma)$, i.e. the number of times that $\sigma''$ appears in the direct sum decomposition of the tensor product of $\sigma$ and $\sigma'$. This can be interpreted as $\hat{\Gamma}_{\sigma,m,n}$ adding $\sigma_m$ to the left vector and $\bar \sigma_n$ to the right vector. Thus, $\hat{\Gamma}_{\sigma,m,n}(\mathbf{x},i)$ on a gauge-invariant state creates net flux $\sigma_m$ at site $\mathbf{x}$ and $\bar{\sigma}_n$ at the site $(\mathbf{x+i})$. Applying $\hat{\Gamma}_{\sigma,m,n}^\dagger$ to the same link annihilates the flux and restores gauge symmetry. Indeed, we can check in Appendix \ref{app:correctability} that $\hat{\Gamma}_\sigma,\hat{\Gamma}_{\sigma'}$ on the same link satisfies the KL condition, 
\begin{align}\label{eq:1-link-correctable}
    \hat{P}_{\mathrm{inv}}{\hat{\Gamma}_\sigma}^\dagger \hat{\Gamma}_{\sigma'}\hat{P}_{\mathrm{inv}}= \delta_{\sigma,\sigma'}\hat{P}_{\mathrm{inv}},
\end{align}
Thus $\hat{\Gamma}_\sigma$-type errors are correctable. Another way to obtain the conclusion is through the code distance $d_{\mathrm{code}}$: for a QEC with code distance $d_{\mathrm{code}}$, errors involving up to $[(d_{\mathrm{code}}-1)/2]$ qudits are guaranteed to be correctable \cite{nielsen_chuang_2010}. In our case,  to make a logical error with $\hat{\Gamma}_\sigma$-type operators, one needs at least four links of a plaquette to form a Wilson loop, thus the code distance is 4. This makes any $[(4-1)/2]=1$  $ \hat{\Gamma}_\sigma$-type error correctable.

The errors induced by the $\hat{L}_{g}(\mathbf{x},i)$ operator are not detectable if $g$ is the Abelian center of the group as it commutes with the $\hat{P}_{\mathrm{inv}}$. For $g$ not in the Abelian center, if $\hat{L}_{g}(\mathbf{x},i)$ drifts a state to another gauge invariant state,  we can not detect such errors. On the other hand when the error $\hat{L}_{g}(\mathbf{x},i)$ drifts a state out of the invariant Hilbert space by generating a local charge at site $\mathbf{x}$, we can measure the local charge at $\mathbf{x}$ and detect such errors. However, as $\hat{L}_{g}(\mathbf{x},i)$ does not affect local charges at other sites, it is not possible to diagnose the link that $\hat{L}_{g}(\mathbf{x},i)$ affects, and thus the errors are not correctable. The fact that $\hat{L}_{g}(\mathbf{x},i)$ are not correctable can also be seen by checking the KL condition:

\begin{align}\label{eq:PLLP}
\hat{P}_{\mathrm{inv}}\hat{L}_{g'}^\dagger\hat{L}_{g}\hat{P}_{\mathrm{inv}} =\frac{1}{|G|} \sum_{h \in G}\hat{L}_{h^{-1}g'^{-1}gh} \hat{P}_{\mathrm{inv}}.
\end{align}
The right hand side of \eq{PLLP} differs from \eq{KL} by an operator $\frac{1}{|G|} \sum_{h \in G}\hat{L}_{h^{-1}g'^{-1}gh}$, thus breaking the KL condition.
In the $Z_2$ gauge theory, a $Z_2$-register is a qubit, and the complete basis reduces to the familiar Pauli operators $\{I, \hat{X},
\hat{Z},\hat{X}\hat{Z}\}$, of which $\hat{\Gamma}_\sigma \in \{I, \hat{Z}\}$ are correctable. 

\begin{figure}
    \centering \includegraphics[width=0.9\linewidth]{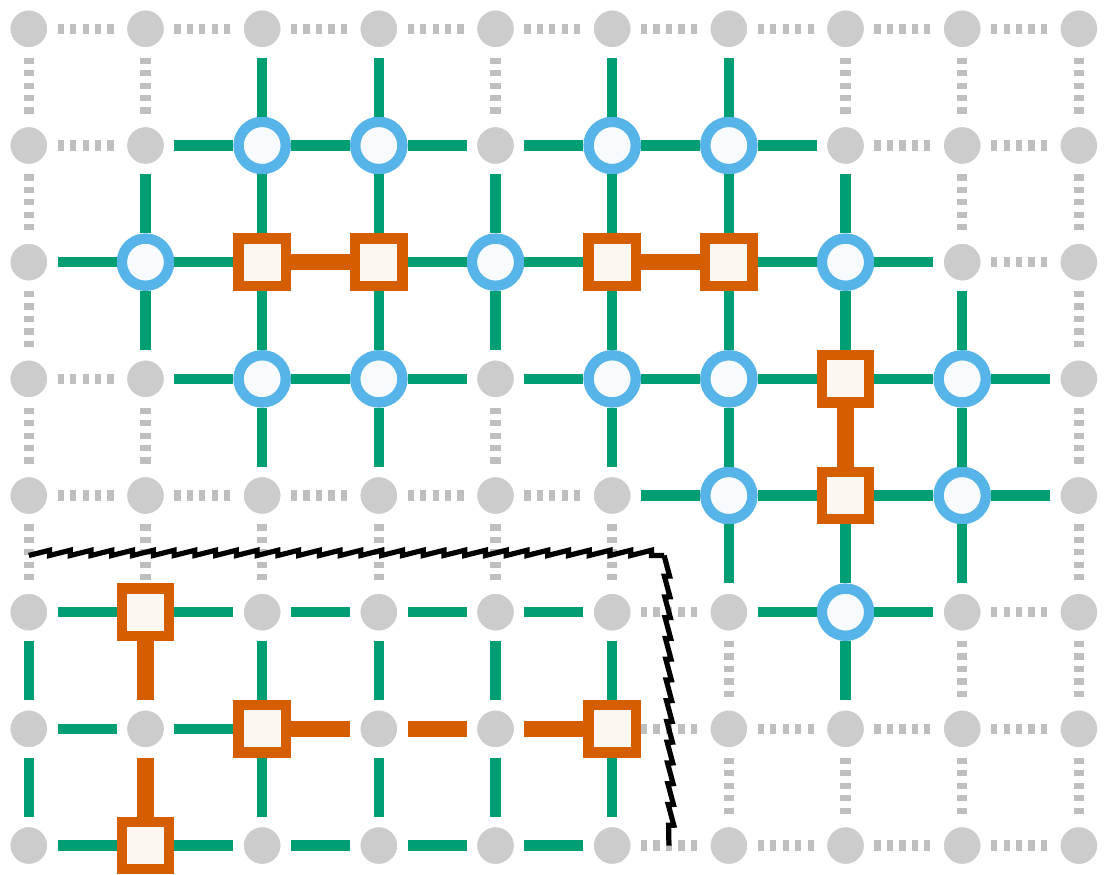}
    \caption{Examples of correctable error distributed on the lattice to satisfy the MED (upper-right) or the local KL condition (lower-left). The orange links are $\Gamma$-type errors, which must break gauge symmetry at the orange squares. The conditions require the green links to be error-less. For the MED condition, the adjacent sites (blue circles) must be charge-neutral.  For the local KL condition, the error links are allowed to extend in a straight line, which is the shortest path to connect a pair of charges.}
    \label{fig:correctable}
\end{figure}

The correctable error set can contain multiple-link $\hat{\Gamma}_\sigma$ errors, as long as the erroneous links are separated enough such that there is no ambiguity about which link causes the gauge violation. The easiest errors to decode are those that show up as pairs of charges in dual representations $\sigma, \bar{\sigma}$ at the two ends of a link, surrounded by neighboring sites all with zero charge. This requires the $\hat{\Gamma}_{\sigma}$ errors to be separated by at least two error-less links (\fig{correctable}), which we call the \textit{minimal effort decoding condition (MED)}. The condition can be relaxed if more complicated classical processing is available. The KL condition requires any product of two error operators $\hat N_i$ and $\hat N_j$ to contain no non-trivial gauge-invariant operators, i.e. Wilson loops. This condition is fulfilled when the number of links affected by $\Gamma_\sigma$ errors along any closed loop of perimeter $C$ is at most $[\frac{C-1}{2}]$, thus eliminating the possibility of error links in $\hat N_i$ and $\hat N_j$ forming loops. Considering only the smallest loops, i.e. plaquettes, the local KL condition allows at most one $\hat{\Gamma}_\sigma$ error per plaquette.

\subsection{The encoding and decoding circuits}\label{sec:circuit}
 
Conceptually, the encoding and decoding of gauge symmetry is easier in the electric basis, where Gauss's law reads as ``zero electric flux at each vertex". 
 For non-Abelian groups, the computation of the net flux requires the group's Clebsch-Gordan coefficients which are not diagonal in the electric basis. We bypass some of the difficulties by doing an ``effective Clebsch-Gordan sum" in the group element basis. We derive in Appendix \ref{app:CG} that the Clebsch-Gordan sum is equivalent to group multiplications in the group element basis. With this, we can encode and decode Gauss's law using the following primitive gates~\cite{Lamm:2019bik} acting on group registers:

\begin{itemize}
    \item Inverse gate: $\mathfrak{U}_{-1}|g\rangle=\left|g^{-1}\right\rangle$,
    \item Left and Right Multiplication gates: $\mathfrak{U}_{\times}^L|g\rangle|U\rangle=|g\rangle|gU\rangle$, $\mathfrak{U}_{\times}^R|g\rangle|U\rangle=|g\rangle| Ug\rangle$,
    \item Fourier transform gate: 
    \begin{equation}
\mathfrak{U}_{F}=\sum\limits_{g,\sigma_{mn}} \sqrt{\frac{d_\sigma}{|G|}} \Gamma^{(\sigma)}_{mn}(g)\ket{\sigma_{mn}}\bra{g}\end{equation} which rotates the magnetic basis to the electric one,
\item Group rotation gates in irreps. This  uses a group register $\ket{U}$ as the control to rotate between the vectors  in the $d_\sigma$-dimensional space: 
\begin{align}\label{eq:Usigma}
    \mathfrak U_{\sigma}&= \sum_{U\in G} (\ket{U}\bra{U}) \otimes (\sum_{m,n=1}^{d_\sigma} \Gamma^{(\sigma)}_{mn}(U) \ket{n}\bra{m})\notag \\&= d_{\sigma}^{-1/2}\sum_{m,n=1}^{d_\sigma} \hat \Gamma_{\sigma, m,n} ^\dagger \otimes (\ket{n}\bra{m})
   \end{align} 
    where the second equality follows from \eq{GM}. For Abelian groups, $\mathfrak U_{\sigma}$ is a one-register phase gate on $\ket{U}$.
\end{itemize}

\begin{figure}
    \centering
    \includegraphics[width=0.3\linewidth]{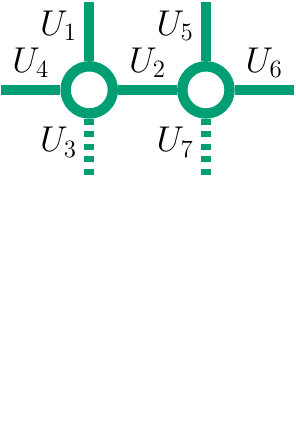}\hspace{-0.3cm}
    \includegraphics[width=0.7\linewidth]{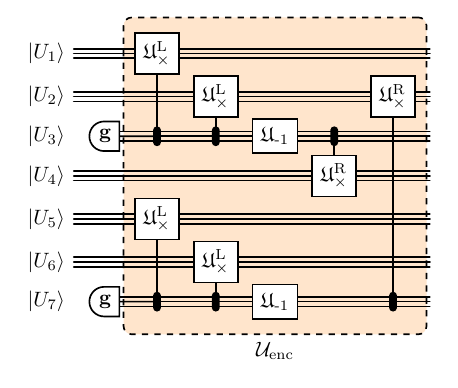}
    \includegraphics[width=\linewidth]{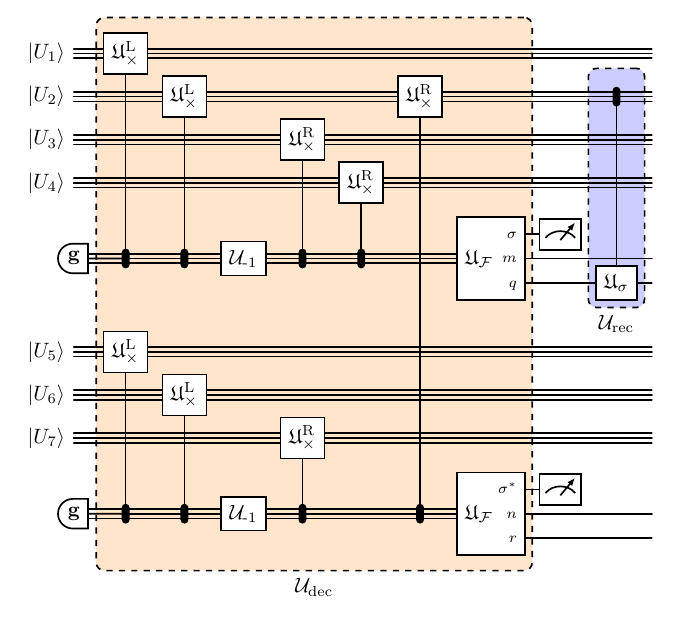}
    \caption{Encoding and decoding circuits for the patch in \fig{maxtree}. (top) Circuit for preparing gauge-invariant states, and (bottom) Gauss's law measurement and recovery circuit, where measuring a pair of charges $\sigma, \bar{\sigma}$ on the two ancillae indicates a gauge-violating error on $\ket {U_2}$.}
    \label{fig:circuit}
\end{figure}

 \fig{circuit} shows an example of encoding and decoding for the patch of 7 gauge links connected to 2 sites in the right corner of \fig{maxtree}. To encode, the links $U_3, U_7$ in the maximal tree are initialized to the color neutral state $\ket{\textbf{g}}\equiv |G|^{-\frac{1}{2}}\sum_{g\in G}\ket{g} $, and then,
 \begin{align}
     \ket{U_1}\ket{U_2}\ket{\mathbf{g}}\ket{U_4}\xrightarrow{\mathfrak{U}_{\mathrm{enc}}} |G|^{-\frac{1}{2}}\sum_{g\in G} \ket{gU_1}\ket{gU_2}\ket{g^{-1}}\ket{U_4g^{-1}}
 \end{align}
 One can easily check the gauge invariance \eq{Gauss} for the two sites after the circuit $\mathcal{U}_{\mathrm{enc}}$ in \fig{circuit}: 

 \begin{align}
     &\hat{T}_{h}\,|G|^{-\frac{1}{2}}\sum_{g\in G} \ket{gU_1}\ket{gU_2}\ket{g^{-1}}\ket{U_4g^{-1}}\notag\\
     =&|G|^{-\frac{1}{2}}\sum_{g'=hg\in G} \ket{g'U_1}\ket{g'U_2}\ket{g'^{-1}}\ket{U_4g'^{-1}}
 \end{align}
In the electric basis, the circuit computes the net flux of $U_1,U_2,U_4$ and stores it into the tree-link $U_3$, thus making the total net flux at the vertex to be zero. The same process is repeated for every vertex except $\mathbf{x_0}$. Sites in the same branch of the maximal tree should be processed in sequence, from the top of the branch to the root, and different branches can be encoded in parallel.

The above procedure initializes the state to the gauge invariant one $\ket{\Psi_{\mathrm{inv}}}$. During the noisy process in \fig{concept}, suppose an error $\hat{\Gamma}_{\sigma,m,n} (\mathbf{x},i)$  turns $\ket{\Psi_{\mathrm{inv}}}$ into $\hat{\Gamma}_{\sigma,m,n} (\mathbf{x},i)\ket{\Psi_{\mathrm{inv}}}$, creating a net  flux of $\sigma_m$ at $\mathbf{x}$ and $\bar{\sigma}_n$ at $\mathbf{x+i}$.
We can decode this error pattern with the circuit $\mathcal{U}_\mathrm{dec}$, which adds the electric fluxes coherently to the ancillary registers originally initialized to color neutral states. Then in the decoding  (derived in Appendix~\ref{app:circuit}),
\begin{align} \label{eq:detection}
    \hat{\Gamma}&_{\sigma,m,n}\ket{\Psi_{\mathrm{inv}}}\otimes \ket{\textbf{g}} \otimes  \ket{\textbf{g}} \xrightarrow{\mathcal{U}_\mathrm{dec}} \notag\\
   &\frac{1}{d_\sigma} \sum_{q,r}^{d_\sigma}\hat{\Gamma}_{\sigma,q,r}\ket{\Psi_{\mathrm{inv}}}\otimes  \ket{\sigma_{mq}(\mathbf{x})}\otimes  \ket{\bar{\sigma}_{nr}(\mathbf{x+i})}.
\end{align}
For non-Abelian groups, $\mathcal{U}_\mathrm{dec}$ preserves the quantum numbers $\sigma, \bar{\sigma}$, but $m,n$ can change into any integer in $[1, d_\sigma]$, and the corresponding new quantum numbers $q,r$ are entangled with the ancillae. 

To correct gauge-violating errors, the representation of the ancillary register at every site needs to be measured. This requires the quantum numbers $\sigma, m,n$ in the Fourier basis to be stored in different qubits, and thus one can measure $\sigma$ without collapsing $m,n$.  Measuring $\ket{\sigma}$ in the trivial representation confirms the local gauge symmetry. Measuring any other $\ket{\sigma}$ indicates a gauge-violating error, and projects the gauge fields to a subspace orthogonal to $\mathcal{H}_{\mathrm{inv}}$ in $\mathcal{H}_{\mathrm{gauge}}$. In this example, a pair of non-trivial representations $\ket{\sigma}, \ket{\bar{\sigma}}$ on the two ends of a link are measured, which indicates a $\hat \Gamma_\sigma$ error on the link, and the quantum numbers $\ket{m}\ket{q},\ket{n}\ket{r}$ of the ancillae are entangled with the gauge field state. The recovery can be performed with $\mathfrak U_\sigma$ in \eq{Usigma}:
  \begin{align}
      &\frac{1}{d_\sigma} \sum_{q,r}^{d_\sigma}\hat{\Gamma}_{\sigma,q,r}\ket{\Psi_{\mathrm{inv}}}\otimes  (\ket{m}\ket{q})\otimes  (\ket{n}\ket{r})\xrightarrow{\mathcal{U}_\mathrm{rec}} \notag\\
       &\frac{1}{d_\sigma^{3/2} }\sum_{q,k,r}^{d_\sigma}\hat{\Gamma}_{\sigma,q,k}^\dagger \hat{\Gamma}_{\sigma,q,r}\ket{\Psi_{\mathrm{inv}}}\otimes  (\ket{m}\ket{k})\otimes  (\ket{n}\ket{r})\notag\\ &=\ket{\Psi_{\mathrm{inv}}}\otimes  \bigg[\frac{1}{d_\sigma^{1/2} }\sum_{r}^{d_\sigma}(\ket{m}\ket{r})\otimes  (\ket{n}\ket{r})\bigg],
  \end{align}
  where we have used the unitarity of representations,
\begin{equation}\label{eq:unitarity}
    d_\sigma^{-1}\sum_{q=1}^{d_\sigma}\hat{\Gamma}_{\sigma,q,k}^\dagger\hat{\Gamma}_{\sigma,q,r}=\delta_{k,r}.
\end{equation}
Thus $\ket{\Psi_{\mathrm{inv}}}$ is recovered and unentangled with ancillae. At this point, since the quantum information in the ancillae is irrelevant to the gauge system, one can recycle the ancillae and re-initialize them to $\ket{\mathbf{g}}$ for the next round of error detection and correction.

\section{Comparison of quantum fidelities}\label{sec:fidelity}

\subsection{Group-agnostic thresholds}
LGT states can be stored in a set of $G$-registers in both the gauge-fixed and gauge-redundant digitization. The number of links needed is $N_L-N_T$ for the former and $N_L$ for the latter. We further assume that errors on different registers are independent. Suppose the total rate for any error on a single register is $\epsilon$. Then the quantum fidelities are lower bounded by the probability that no errors happen on any register:
\begin{align}\label{eq:F_epsilon}
    F_{\mathrm{fixed}} &\geq (1-\epsilon)^{N_L-N_T}\\
    F_{\mathrm{red}} &\geq (1-\epsilon)^{N_L}
\end{align}
 With the gauge redundancy, certain configuration of $\hat \Gamma_{\sigma}$ errors are correctable via gauge symmetry checking and restoration. Suppose the correctable type error rate on a single register is $\epsilon_c$.
The lower bound for the fidelity after gauge-symmetry restoration is the probability that no error or only correctable errors happen:
\begin{align}\label{eq:restored}
    F_{\mathrm{restored}} &\geq \sum_{n=0}^{N_L} Q_n \epsilon_c^n (1-\epsilon)^{N_L-n}=(1-\epsilon)^{N_L} \Xi 
\end{align}
where $Q_n$ is the number of ways to arrange $n $ links with correctable errors in the lattice, such that the correctability condition is still satisfied, and $\Xi $ is the factor by which the gauge-symmetry restoration amplifies the fidelity:  
\begin{align}\label{eq:Xi}
    \Xi = \sum_{n=0}^{N_L}Q_n z^n,\quad z=\frac{\epsilon_c}{1-\epsilon}
\end{align}

Comparing these two bounds \eq{restored} and \eq{F_epsilon}, the condition for the gauge-symmetry restoration to give a higher fidelity than gauge fixing is
\begin{align}\label{eq:threshold}
    \Xi^{1/N_T} > (1-\epsilon)^{-1}.
\end{align}
This formula can be interpreted as an error threshold: the gauge redundancy increases the code distance of $\Gamma_\sigma$-type errors from 1 to 4. For the fidelity to increase with the increased code distance due to gauge redundancy, $\epsilon$ has to be below certain threshold.

\begin{figure}
    \centering
        \includegraphics[width=0.85\linewidth]{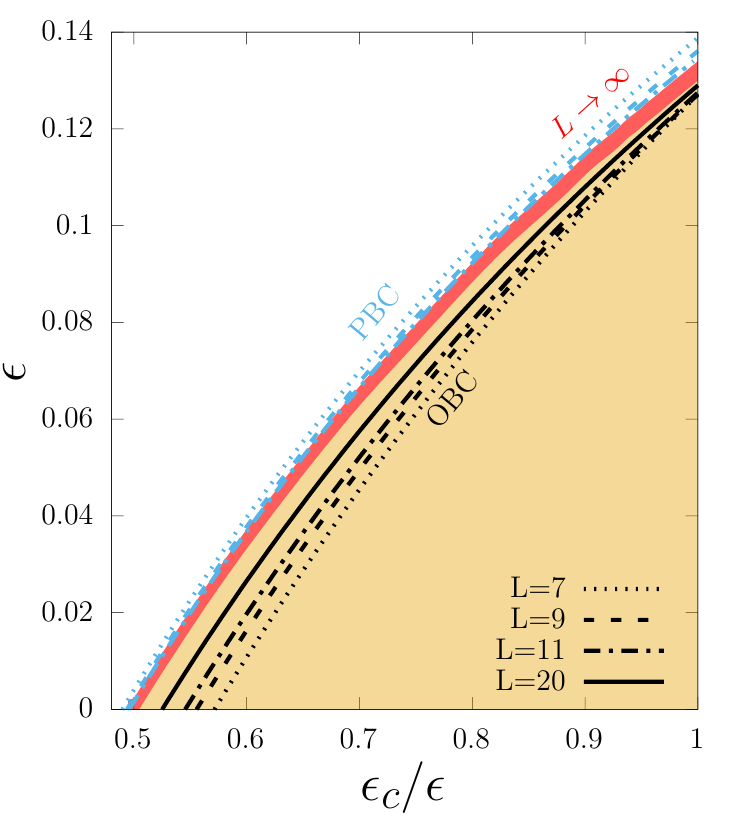}
       \includegraphics[width=0.85\linewidth]{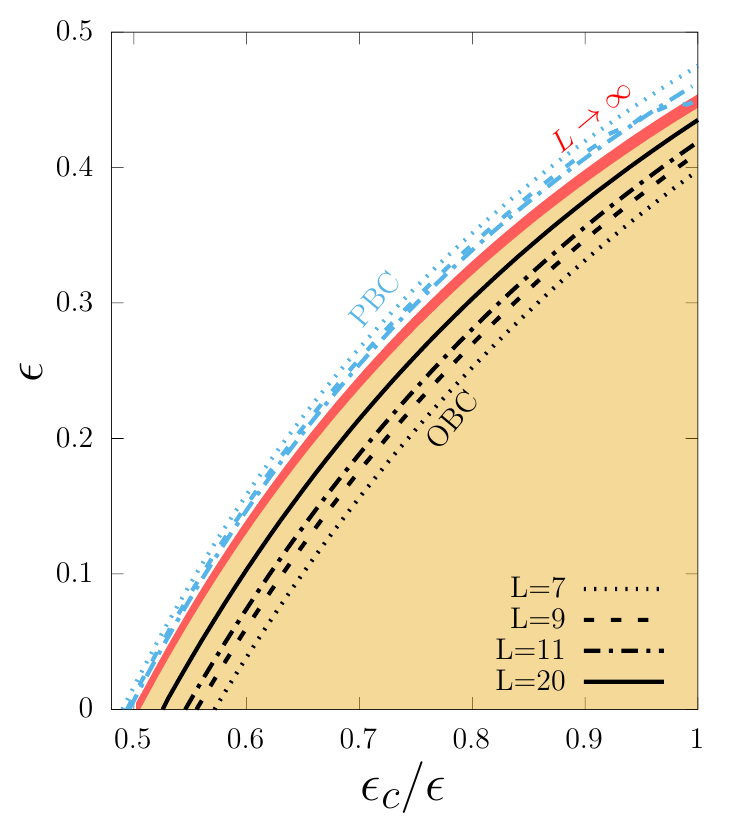}
      \caption{Error threshold lines in the $\epsilon -\epsilon_c/\epsilon$ plane below which $F_{\mathrm {restored}}>F_{\mathrm{fixed}}$, indicating that using gauge redundancy to detect and correct the errors is advantageous, for 2d lattices for (top) MED condition and (bottom) local KL condition. The shaded region is the infinite volume limit. }         \label{fig:2dthresholds} 
\end{figure}

\begin{figure}
  \centering
         \includegraphics[width=0.85\linewidth]{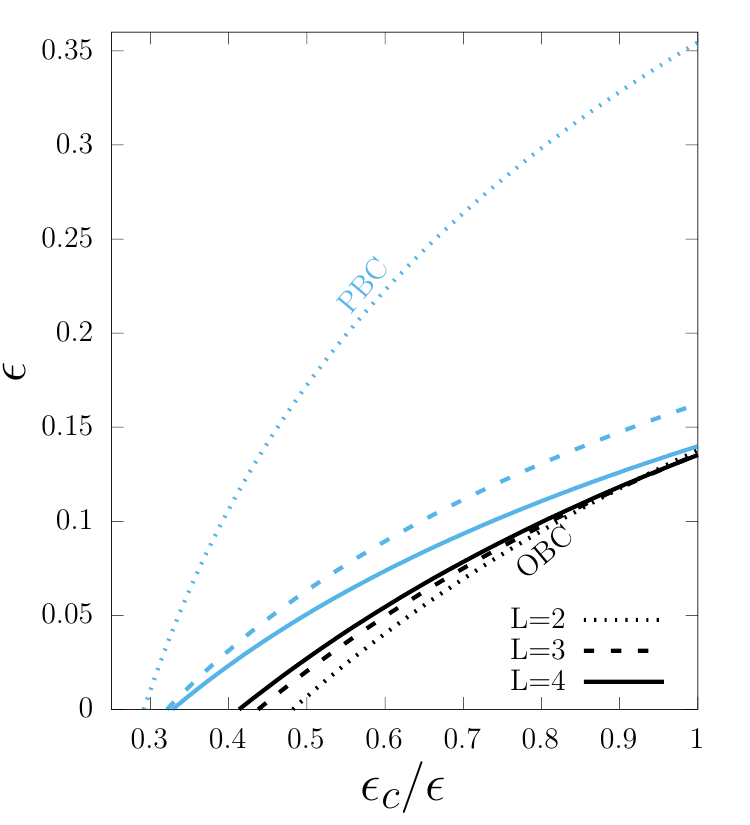}
         \includegraphics[width=0.85\linewidth]{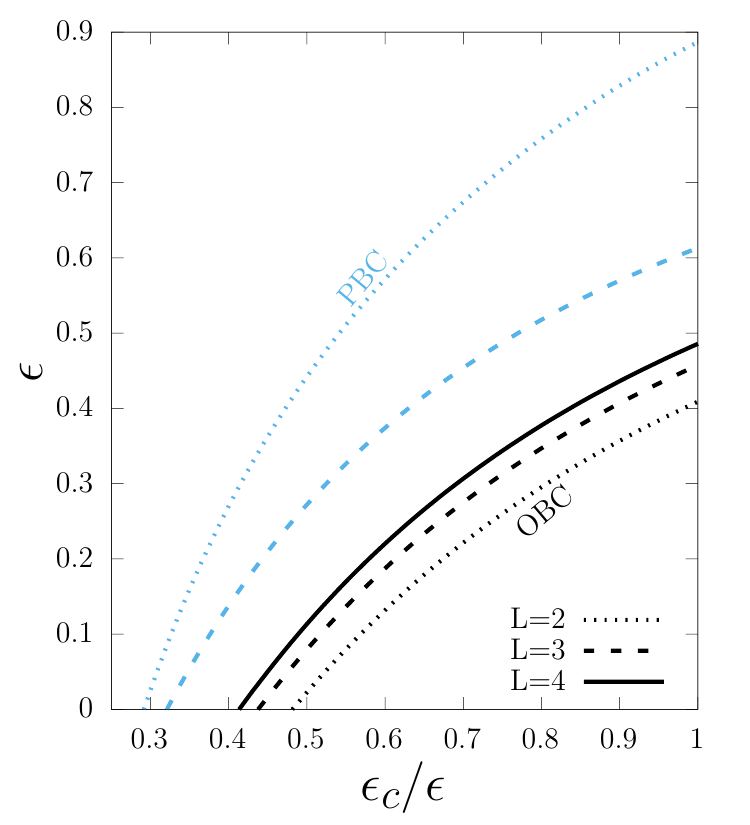}
 \caption{Error threshold lines in the $\epsilon -\epsilon_c/\epsilon$ plane below which $F_{\mathrm {restored}}>F_{\mathrm{fixed}}$, indicating that using gauge redundancy to detect and correct the errors is advantageous, for 3d lattices for (top) MED condition and (bottom) local KL condition.}   
\label{fig:3dthresholds}    
\end{figure}

The $Q_n$'s are related to the independence polynomials of graphs (see Appendix~\ref{app: graph} for details). We  compute $Q_n$ for a variety of $L$ using the Python library \verb|hobj| ~\cite{Butera:2014nva} up to $L=20$ and $L=11$ for 2d lattices with OBC and PBC, respectively, for both error-correction conditions.  For 3d lattices, we are able to reach $L=4$ for both OBC and PBC with MED condition, while only $L=3$ for PBC with the KL condition due to limited computing resources.  These $Q_n$ are then used to compute the threshold: the $\epsilon$ where the two sides of \eq{threshold} are equal at a given $\epsilon_c/\epsilon$. These are shown in black (blue) lines using OBC (PBC) in Fig.~\ref{fig:2dthresholds} and Fig.~\ref{fig:3dthresholds} for 2d and 3d lattices, respectively. \eq{threshold} is satisfied in the parameter region below the lines, indicating that the gauge symmetry restoration provides higher fidelity than the gauge fixed case. 

We observe two features from Fig.~\ref{fig:2dthresholds}-\ref{fig:3dthresholds}. First,the parameter regions to satisfy \eq{threshold} only exist when $\epsilon_c/\epsilon\gtrsim1/d$, which is consistent with the following analysis. Expanding \eq{threshold} for small $\epsilon$, with $Q_0=1, Q_1=N_L$, we have:
\begin{align}
    1+\frac{N_L}{N_T} \epsilon_c+ O(\epsilon_c\epsilon)> 1+\epsilon+O(\epsilon^2).
\end{align}
Thus, the correctable fraction of the total error rate for a single link, $\epsilon_c/\epsilon$, must satisfy
\begin{align}
    \frac{\epsilon_c}{\epsilon}>\frac{N_T}{N_L}\approx \frac{1}{d},
\end{align}
with $\frac{N_T}{N_L}= \frac{1}{d}$ in the infinite volume limit as seen in \tab{DOF}. Second, in all the cases, the threshold $\epsilon$ increases as $\epsilon_c/\epsilon$ increases. This is consistent with the qualitative reasoning that when a larger portion of errors are correctable,  larger values of $\epsilon$ could be allowed while the redundancy still enables more correction than the errors it introduces.

Based on the error thresholds obtained at different $L$, we aim to extrapolate to $L\rightarrow\infty$ limit.  
We find that near $\epsilon=0$, the threshold is a monotonic function of $L$ in either boundary condition and approaches the same limit.  In contrast, near $\epsilon_c/\epsilon=1$ for 2d lattices, non-monotonic behavior is observed for small lattices with $L\le 6$ due to finite volume effects. By restricting to the results for $L>6$ at 2d lattice, we perform both quadratic and exponential extrapolations to obtain the threshold $\epsilon$ in the $L\rightarrow\infty$ limit which can be found in Fig.~\ref{fig:2dthresholds} (red line with the thickness quantifying error bar on extrapolations).
For 3d lattice, as computing resources limit the calculations to only $L<5$, the extrapolation to $L\rightarrow \infty$ cannot be reliably performed. For MED condition, the threshold curve with OBC near $\epsilon_c/\epsilon=1$ exhibits a larger deviation from the PBC results at larger $L$, suggesting that $L<5$ is far from the infinite limit. For the KL condition, calculations are limited to only $L=2,3$ with PBC, also precluding an extrapolation to infinity limit. Despite this, we expect the thresholds for $L\rightarrow\infty$ to be roughly bounded by the largest $L$ results for OBC and PBC.

\subsection{\texorpdfstring{$\epsilon, \epsilon_c$}{e,ec} from error models}
$\hat N_j$ on a $G$-register can always be decomposed as
\begin{align}
    \hat{N}_j=\sum_{g,\sigma} c_{j,g,\sigma} \hat{L}_g \hat{\Gamma}_\sigma.
\end{align}
The trace-preserving feature $\sum_j \hat{N}_j^\dagger \hat{N}_j=\mathbb{1}$ requires $\sum_{j,\sigma, g} |c_{j,g,\sigma} |^2=1$. When the noise channel is diagonal in this basis, i.e. for each $\hat{N}_j$ there is only one non-zero $c_{j,g,\sigma}$, such that we can relabel $c_{j,g,\sigma}$ to $c_{g, \sigma}$. The one-register noise channel can be written as
\begin{align}
   \mathcal{N}(\rho)=\sum_{ g,\sigma} |c_{g,\sigma} |^2 \hat{L}_g \hat{\Gamma}_\sigma \rho (\hat{L}_g \hat{\Gamma}_\sigma)^\dagger.
\end{align}
The one register total error probability is 
\begin{equation}
    \epsilon=\sum_{g\neq \mathbb{1}, \sigma \neq \mathbb{1}} |c_{g,\sigma} |^2=1-|c_{\mathbb{1},\mathbb{1}} |^2. 
\end{equation}
where $g\neq\mathbb{1}$ and $\sigma \neq \mathbb{1}$ mean, respectively, that the group element cannot be the identity, and the irreducible representation cannot be the trivial one. With this, the correctable error probability is
\begin{equation}
    \epsilon_c=\sum_{\sigma \neq \mathbb{1}} |c_{\mathbb{1},\sigma} |^2
\end{equation}

We consider $Z_N$ theory, where each $G$-register is a qudit with $N$ states \cite{Gustafson:2023swx}. We represent the computational basis of the qudit as $\ket{n}$, where $n$ is an integer in $[0,N-1]$. In the computational basis, define the clock shift operators
\begin{equation}
    \hat{\chi}=\sum_{n=0}^{N-1} \ket{n}\bra{{(n+1)\text{ mod }N}}
\end{equation} and the phase shift operators 
\begin{equation}
    \hat{\nu}=\sum_{n=0}^{N-1} e^{2 \pi i n/N}\ket{n}\bra{n}.
\end{equation}  For the group $Z_N$, both group elements $g$ and group representations $\sigma$ can be mapped to integers $[0,N-1]$. Thus we can choose to have the computational basis as either the group element or the electric basis. If we choose the group element basis, $\hat{L}_g$ are clock shifts 
\begin{equation}
    \hat{L}_g=(\hat{\chi})^g=\sum_{n=0}^{N-1} \ket{n}\bra{{(n+g)\text{ mod }N}},
\end{equation}
which preserves the gauge symmetry, and thus cannot be diagnosed by Gauss's law. The representation $\hat{\Gamma}_\sigma$ operators are phase shifts
 \begin{equation}
    \hat{\Gamma}_\sigma=(\hat{\nu})^\sigma=\sum_{n=0}^{N-1} e^{2 \pi i n\sigma/N}\ket{n}\bra{n}.
\end{equation}
They break gauge symmetry unless they form closed Wilson loops. 

If we choose the electric basis as the computational basis instead, as the phase shift is mapped to the $\hat{L}_g$ operator as $\hat{L}_g=(\hat{\nu})^g$, thus the phase shift errors would preserve the symmetry. While the clock shift is mapped to $\hat{\Gamma}_\sigma$ as 
$\hat{\Gamma}_\sigma=(\hat{\chi})^\sigma$, making the clock shift errors correctable.

Consider the error channel where single clock and phase shifts occur independently with probabilities $p_\chi, p_\nu$ :
\begin{align}
    \mathcal{N}(\rho)=&(1-p_\chi)(1-p_\nu)\rho \notag\\
    &+  \frac{(1-p_\chi)p_\nu}{N-1}\sum_{i=1}^{N-1} (\hat{\nu})^i \rho{(\hat{\nu})^i}^\dagger\notag\\
    &+  \frac{(1-p_\nu)p_\chi}{N-1}\sum_{j=1}^{N-1} (\hat{\chi})^j \rho{(\hat{\chi})^j}^\dagger\notag\\
    &+\frac{p_\nu p_\chi}{(N-1)^2}\sum_{i,j=1}^{N-1}(\hat{\nu})^i (\hat{\chi})^j \rho{(\hat{\chi})^j}^\dagger{(\hat{\nu})^i}^\dagger
\end{align}
The one register total error probability is 
\begin{equation}
    \epsilon= 1- (1-p_\chi)(1-p_\nu)
\end{equation}
In the group element basis, $(\hat{\nu})^\sigma $ are correctable, and thus
\begin{equation}
    \epsilon_c=(1-p_\chi)p_\nu.
\end{equation}
In the electric basis, $(\hat{\chi})^\sigma$ are correctable, 
\begin{equation}
    \epsilon_c=(1-p_\nu)p_\chi.
\end{equation}
From this, we observe that the relative error probabilities of physical hardware can prefer different encodings. If the phase errors are more frequent than the clock errors ($p_\nu>p_\chi$), encoding in the group element basis will help to stay below the error thresholds with the total $\epsilon$ reasonably small.

The correctability of phase errors can be generalized to non-Abelian groups, if one chooses the group element basis encoding. This is because all operators diagonal in the group element basis can be written as linear combinations of $\hat{\Gamma}_\sigma$. For example, in \tab{Q8}, we list the correctable errors for the 3-qubit encoding of the quaternion group $\mathbb Q_8=\{(-1)^a \mathbf{i}^b \mathbf{j}^c\}$, where $a,b,c=0$ or $1$~\cite{Gustafson:2023swx}. It has four 1d representations and one 2d representation. All the Pauli-Z operators are correctable, as $\hat{Z}_b, \hat{Z}_c$ correspond to two 1-d representation operators, and $\hat{Z}_a$ is a linear combination of 2-d representation operators:
\begin{equation}
    \hat{Z}_a=\frac{1}{1+i}(\hat\Gamma_{2,1,1}+\hat\Gamma_{2,1,2}-i \hat\Gamma_{2,2,1}+i \hat\Gamma_{2,2,2}).
\end{equation}

\begin{table}[]
    \centering
    \begin{tabular}{c c|c c}
    \hline
       \multicolumn{2}{c|}{1-d representations}  &  \multicolumn{2}{|c}{2-d representation} \\
       \hline 
       $\hat\Gamma_{\mathrm{triv}}$ & $\hat I$ & $\hat\Gamma_{2,1,1}$ & $\hat Z_a \hat {Z_b}^{1/2} \frac{1+\hat Z_c}{2}$\\
        $\hat\Gamma_{\mathrm{i-ker}}$ & $\hat Z_c$ & $\hat\Gamma_{2,1,2}$ & $\hat Z_a \hat {Z_b}^{1/2} \frac{1-\hat Z_c}{2}$\\
        $\hat\Gamma_{\mathrm{j-ker}}$ & $\hat Z_b$ & $\hat\Gamma_{2,2,1}$ & $-\hat Z_a \hat {Z_b}^{3/2} \frac{1-\hat Z_c}{2}$\\
         $\hat\Gamma_{\mathrm{k-ker}}$ & $\hat Z_b\hat Z_c$& $\hat\Gamma_{2,2,2}$ & $\hat Z_a \hat {Z_b}^{3/2} \frac{1+\hat Z_c}{2}$\\
       \hline
    \end{tabular}
\caption{Correctable errors $\hat{\Gamma}_{\sigma}$ for the quaternion group $\mathbb Q_8$ encoded in 3 qubits via $(-1)^a \mathbf{i}^b \mathbf{j}^c \to | abc \rangle$.}
    \label{tab:Q8}
\end{table}
Thus, for the error model that allows one of the three qubits to have a Pauli error,
\begin{align}
    \mathcal{N}(\rho)=& (1-3p_X-3p_Y-3p_Z) \rho  + p_X \sum_{i=a,b,c}\hat{X}_i \rho \hat X_i\notag\\
   &+ p_Y \sum_{i=a,b,c}\hat{Y}_i \rho \hat Y_i+p_Z\sum_{i=a,b,c}\hat{Z}_i \rho \hat Z_i
\end{align}
with the one-register error rate being $3(p_X+p_Y+p_Z)$, the correctable fraction is 
\begin{equation}
    \frac{\epsilon_c}{\epsilon}=\frac{p_Z}{p_X+p_Y+p_Z}.
\end{equation}
If the qubits have more phase errors and fewer bit-flip errors, encoding in the group element basis is preferable.
\section{Conclusions}
In this work, we explore how the natural gauge redundancy of lattice gauge theories can become a tool to create partial error correction codes for quantum simulations. For generic groups, treating the gauge transformations in lattice gauge theories as stabilizers and the gauge-invariant subspace as the code space, we identify the correctable errors in one lattice link, as well as the MED and KL conditions for multiple link errors to remain correctable. We construct the quantum circuits to prepare gauge invariant states including redundant degrees of freedoms, as well as to detect and recover the correctable errors. 

We calculate the error-rate thresholds below which keeping gauge redundancy is preferable to gauge fixing for error-correcting purposes, under both the MED and KL conditions. We do this by comparing the quantum fidelities when errors exist, which is the probability that either no error happens or only  correctable errors happen. The thresholds depend on the correctable fraction of the error rate in a single link, $\epsilon_c/\epsilon$. We find a simple analytic relation at leading order in $\epsilon$ for when gauge redundancy is advantageous: $\epsilon_c/\epsilon \geq 1/d$ for a $d$ dimensional lattice. Numerical results demonstrate that this relation is robust for realistic error rates. Thus for quantum devices where error correction is possible, $\epsilon\ll 1$, our results can be a guidance to design digitizations. 
To provide examples of how one may apply such thresholds, we obtain $\epsilon_c/\epsilon$ explicitly for the discrete Abelian group $\mathbb{Z}_N$ and the non-Abelian group $\mathbb Q_8$ assuming certain reasonable error models.

Consideration of similar error thresholds should be generalized to other digitization schemes and error models. The quantitative values of the thresholds found in this paper can vary depending on the computational tasks and the hardware. Future research directions should also take into account non-diagonal error channels, quantum architectures and inclusion of fermions on the lattice sites. 

\begin{acknowledgments}
The authors thank Mario Pernici for his invaluable assistance with \verb|hobj|.
This work is supported by the Department of Energy through the Fermilab QuantiSED program in the area of ``Intersections of QIS and Theoretical Particle Physics". Fermilab is operated by Fermi Research Alliance, LLC under contract number DE-AC02-07CH11359 with the United States Department of Energy. Y.-Y. L is supported by the NSF of China through Grant No. 12305107, 12247103.
\end{acknowledgments}

\bibliography{refs}% Produces the bibliography via BibTeX.

\appendix
\section{Counting gauge orbits} \label{app:orbits}
Let $\mathcal{U} $ be the set of configurations and $\mathcal{G}$ the group of gauge transformations on the entire lattice, with $|\mathcal{U}|=|G|^{N_L}$ and $|\mathcal{G}|=|G|^{N_V}$.   In the magnetic basis, $\hat{P}_{\rm{inv}}\ket{ U}$ is the uniform superposition of all gauge transformations of $\ket{U}$. Thus the number of states in the basis $\hat{P}_{\rm{inv}}\mathcal{U}$ is the number of configurations in $\mathcal{U}$ that are not equivalent under gauge transformations:
 \begin{equation}
     \mathrm{dim} \mathcal{H}_{\mathrm{inv}}=|\mathcal{U}/\mathcal{G}|
 \end{equation}
 where $|\mathcal{U}/\mathcal{G}|$ is the number of orbits in $\mathcal{U}$ under $\mathcal{G}$. 
According to Burnside's counting theorem, 
\begin{align}
    |\mathcal{U/G}|=\frac{1}{|\mathcal{G}|} \sum_{ T \in \mathcal{G}} |\mathcal{U}^T|=\frac{1}{|\mathcal{G}|} \sum_{ \ket{U} \in \mathcal{U}} |\mathcal{G}_{\ket{U}}|,
\end{align}
where $\mathcal{U}^T$ is the subset of elements invariant under $T\in \mathcal{G}$,
\begin{align}
   \mathcal{U}^T= \{\ket{U} \in \mathcal{U}: T \ket{U}=\ket{U}\},
\end{align}
 and $\mathcal{G}_{\ket{U}}$ is the stabilizer group for the element $\ket{U}$,
 \begin{align}
     \mathcal{G}_{\ket{U}}=\{T \in \mathcal{G}: T \ket{U}=\ket{U}\}.
 \end{align}
If $T=\prod _\mathbf{x} \hat{T}_{h_\mathbf{x}}$ then with $\ket{U}=\prod _{\mathbf{x},i} \ket{g_{\mathbf{x},i}}$, there must be
\begin{align}
    h_{\mathbf{x+i}}=g_{\mathbf{x},i}^{-1} h_{\mathbf{x}} g_{\mathbf{x},i},
\end{align}
which means once $h_{\mathbf{x}_0} $ at a certain site $\mathbf{x}_0$ is fixed, there is at most one $T\in \mathcal{G}$ to keep $\ket{U}$ invariant. For $h_{\mathbf{x}_0} \in H$ the Abelian center of $G$, this is the global transformation with $h_{\mathbf{x}_0} $. Therefore $|H|\leq |\mathcal{G}_{\ket{U}}|\leq  |G|$ and 
\begin{align}
     |\mathcal{U}||H|\leq \sum_{ \ket{U} \in \mathcal{U}} |\mathcal{G}_{\ket{U}}|\leq  |\mathcal{U}||G|.
\end{align}
given that $|\mathcal{U}|=|G|^{N_L},|\mathcal{G}|=|G|^{N_V}$, we find
\begin{align}
     |H||G|^{N_L-N_V}\leq |\mathcal{U/G}| \leq |G|^{N_L-N_V+1}.
\end{align}
For Abelian groups, $H=G$ and $ |\mathcal{U/G}|=|G|^{N_L-N_V+1}=|G|^{N_L-N_T}$.  For non-Abelian $G$, the ``$\leq$" are both ``$<$".

\begin{widetext}

\section{Correctability of \texorpdfstring{$\hat{\Gamma}_\sigma$}{Gs}} \label{app:correctability}
In the rest of the appendices, $g,h$ are group elements, $\sigma,\tau$ are irreducible representations (irreps), and $k,l,m,n,q,r,s,t,u$ are the matrix elements of representations.To check \eq{1-link-correctable}, we will compute in the magnetic basis
\begin{equation}
 \hat{P}_0(\mathbf{x})\hat{\Gamma}_{\sigma',s, q}^\dagger(\mathbf{x},i)\hat{\Gamma}_{\sigma,m, n}(\mathbf{x},i)\hat{P}_0(\mathbf{x}) = \frac{1}{|G|^2} \sum_{g,h\in G}  \bigg (\vcenter{\hbox{\begin{tikzpicture}
    % Draw the cross
    \draw (0,0.6) -- (0,-0.6); % Vertical line
    \draw (-0.6,0) -- (0.6,0); % Horizontal line
    
    % Add labels
    \node[above] at (0,0.6) {$\hat{L}_g\hat{L}_h$};
    \node[below] at (0,-0.6) {$\hat{R}_{g^{-1}}\hat{R}_{h^{-1}}$};
    \node[left] at (-0.6,0) {$\hat{R}_{g^{-1}}\hat{R}_{h^{-1}}$};
    \node[right] at (0.6,0) {$\hat{L}_g\hat{\Gamma}^\dagger_{\sigma',s,q}\hat{\Gamma}_{\sigma,m,n}\hat{L}_h$};
    
    % Add the site label
    \node[draw=black, fill=white, circle, inner sep=2pt] at (0,0) {$\mathbf{x}$};
  \end{tikzpicture}}} \bigg ),
\end{equation}
For the operators on the link $\mathbf{x},i$,
\begin{align}
    \hat{L}_g\hat{\Gamma}^\dagger_{\sigma',s,q}\hat{\Gamma}_{\sigma,m,n} \hat{L}_h&= \sum_{U}\ket{gU}\bra{U} \sqrt{d_\sigma d_{\sigma'}} \Gamma^{(\sigma')}_{sq}(
U) \Gamma^{(\sigma)}_{mn}(
U)^\ast \hat{L}_h\notag\\
&=\sqrt{d_\sigma} \bigg[\sum_{U,l}\Gamma^{(\sigma')}_{sl}(
g^{-1})\Gamma^{(\sigma')}_{lq}(gU)\bigg]\bigg[\sum_{k}\Gamma^{(\sigma)}_{mk}(
g^{-1})\Gamma^{(\sigma)}_{kn}(gU)\bigg]^\ast\ket{gU}\bra{U}  \hat{L}_h\notag\\
&= \sum_{k,l}\Gamma^{(\sigma')}_{sl}(
g^{-1})\Gamma^{(\sigma)}_{mk}(
g^{-1})^\ast \hat{\Gamma}_{\sigma',l,q}^\dagger\hat{\Gamma}_{\sigma,k,n}\hat{L}_g \hat{L}_h.
\end{align}
Thus, let $h'=gh$,
\begin{equation}
\hat{P}_0(\mathbf{x})\hat{\Gamma}_{\sigma',s, q}^\dagger(\mathbf{x},i)\hat{\Gamma}_{\sigma,m, n}(\mathbf{x},i)\hat{P}_0(\mathbf{x}) = \frac{1}{|G|^2}\sum_{h'\in G}  \bigg (\vcenter{\hbox{\begin{tikzpicture}
    % Draw the cross
    \draw (0,0.6) -- (0,-0.6); % Vertical line
    \draw (-0.6,0) -- (0.6,0); % Horizontal line
    
    % Add labels
    \node[above] at (0,0.6) {$\hat{L}_{h'}$};
    \node[below] at (0,-0.6) {$\hat{R}_{h'^{-1}}$};
    \node[left] at (-0.6,0) {$\hat{R}_{h'^{-1}}$};
    \node[right] at (0.6,0) {$\sum\limits_{k,l,g}\Gamma^{(\sigma')}_{sl}(
g^{-1})\Gamma^{(\sigma)}_{mk}(
g^{-1})^\ast \hat{\Gamma}_{\sigma',l,q}^\dagger\hat{\Gamma}_{\sigma,k,n}\hat{L}_{h'}$};
    
    % Add the site label
    \node[draw=black, fill=white, circle, inner sep=2pt] at (0,0) {$\mathbf{x}$};
  \end{tikzpicture}}} \bigg ).
\end{equation}
With the orthogonality of matrix elements
\begin{equation}\label{eq:ortho}
    \sum_{g}\Gamma^{(\sigma')}_{sl}(
g^{-1})\Gamma^{(\sigma)}_{mk}(
g^{-1})^\ast =\delta_{\sigma\sigma'}\delta_{ms} \delta_{lk} |G|/d_{\sigma},
\end{equation}
and the unitarity of representations \eq{unitarity}, we get
\begin{equation}
\hat{P}_0(\mathbf{x})\hat{\Gamma}_{\sigma',s, q}^\dagger(\mathbf{x},i)\hat{\Gamma}_{\sigma,m, n}(\mathbf{x},i)\hat{P}_0(\mathbf{x}) = \delta_{\sigma'\sigma}\delta_{sm}\delta_{qn}\hat{P}_0(\mathbf{x}) 
\end{equation}
\section{Derivation of \eq{detection} in the group element basis} \label{app:circuit}

 $\mathcal{U}_\mathrm{dec}$ performs the following for a gauge field state $\ket{\Psi}$:
\begin{align}\label{eq:circuit}
    \ket{\Psi} \otimes (|G|^{-\frac{1}{2}} \sum_{g\in G}\ket{g}) &\xrightarrow{\mathfrak{U}_{\times}^L,\mathfrak{U}_{\times}^L,\mathfrak{U}_{-1},\mathfrak{U}_{\times}^R,\mathfrak{U}_{\times}^R} |G|^{-\frac{1}{2}} \sum_{g\in G} \hat{T}_{g}(\mathbf{x}) \ket{\Psi} \otimes \ket{g^{-1}}\notag\\
   & \xrightarrow{QFT}   \sum_{g\in G} \hat{T}_{g}(\mathbf{x}) \ket{\Psi} \otimes \bigg(\sum_{\sigma,m,n} \frac{\sqrt{d_\sigma}}{|G|}\Gamma^{(\sigma)}_{mn}(g^{-1})\ket{\sigma}\bigg)=\sum_{\sigma,m,n} \hat{P}_{\sigma_{mn}}(\mathbf{x})\ket{\Psi}\otimes \ket{\sigma_{mn}},
   \end{align}
   where 
   \begin{equation}\label{eq:P_sigma_mn}
        \hat{P}_{\sigma_{mn}}(\mathbf{x})\equiv \frac{\sqrt{d_{\sigma}}}{|G|}\sum_{g\in G} \Gamma^{(\sigma)}_{mn}(g^{-1})\hat{T}_{g}(\mathbf{x})
   \end{equation}
   For the trivial representation $\sigma=0$, $\hat{P}_0(\mathbf{x})$
is the projection to the subspace invariant under local gauge transformations and $\hat{P}_0(\mathbf{x}) \ket{\Psi}_{\mathrm{inv}}=\ket{\Psi}_{\mathrm{inv}}$. When a correctable error $\hat{\Gamma}_{\sigma}$ occurs on one gauge register, consider

\begin{equation}\label{eq:cross}
 \hat{P}_{\sigma'_{sq}}(\mathbf{x})\hat{\Gamma}_{\sigma,m, n}(\mathbf{x},i)\hat{P}_0(\mathbf{x}) = \frac{\sqrt{d_{\sigma}}}{|G|^2}\sum_{g,h\in G} \Gamma^{(\sigma')}_{sq}(g^{-1}) \bigg (\vcenter{\hbox{\begin{tikzpicture}
    % Draw the cross
    \draw (0,0.6) -- (0,-0.6); % Vertical line
    \draw (-0.6,0) -- (0.6,0); % Horizontal line
    
    % Add labels
    \node[above] at (0,0.6) {$\hat{L}_g\hat{L}_h$};
    \node[below] at (0,-0.6) {$\hat{R}_{g^{-1}}\hat{R}_{h^{-1}}$};
    \node[left] at (-0.6,0) {$\hat{R}_{g^{-1}}\hat{R}_{h^{-1}}$};
    \node[right] at (0.6,0) {$\hat{L}_g\hat{\Gamma}_{\sigma,m,n}\hat{L}_h$};
    
    % Add the site label
\node[draw=black, fill=white, circle, inner sep=2pt] at (0,0) {$\mathbf{x}$};
  \end{tikzpicture}}} \bigg ),
\end{equation}
use the ``commutation relation" to replace $\hat{L}_g\hat{\Gamma}_\sigma$ on the link $(\mathbf{x},i)$ with 
\begin{equation}
    \hat{L}_g\hat{\Gamma}_{\sigma,m,n}= \sum_{U}\ket{gU}\bra{U} \sqrt{d_\sigma} \Gamma^{(\sigma)}_{mn}(
U)^\ast =\sum_{U}\ket{gU}\bra{U} \sqrt{d_\sigma} [\sum_{k}\Gamma^{(\sigma)}_{mk}(
g^{-1})\Gamma^{(\sigma)}_{kn}(gU)]^\ast= \sum_{k}\Gamma^{(\sigma)}_{mk}(
g^{-1})^\ast \hat{\Gamma}_{\sigma,k,n}\hat{L}_g,
\end{equation}
and let $h'=gh$, \eq{cross} can be simplified as
\begin{equation}
         \hat{P}_{\sigma'_{sq}}(\mathbf{x})\hat{\Gamma}_{\sigma,m,n}(\mathbf{x},i)\hat{P}_0(\mathbf{x}) = \frac{\sqrt{d_{\sigma}}}{|G|^2}\sum_{g} \Gamma^{(\sigma')}_{sq}(g^{-1}) \sum_k\Gamma^{(\sigma)}_{mk}(
g^{-1})^\ast \sum_{h'}\bigg (\vcenter{\hbox{\begin{tikzpicture}
    % Draw the cross
    \draw (0,0.6) -- (0,-0.6); % Vertical line
    \draw (-0.6,0) -- (0.6,0); % Horizontal line
    
    % Add labels
    \node[above] at (0,0.6) {$\hat{L}_{h'}$};
    \node[below] at (0,-0.6) {$\hat{R}_{h'^{-1}}$};
    \node[left] at (-0.6,0) {$\hat{R}_{h'^{-1}}$};
    \node[right] at (0.6,0) {$\hat{\Gamma}_{\sigma,k,n}\hat{L}_{h'}$};
    
    % Add the site label
\node[draw=black, fill=white, circle, inner sep=2pt] at (0,0) {$\mathbf{x}$};
  \end{tikzpicture}}} \bigg ).
\end{equation}
Using the orthogonality \eq{ortho}, we get
\begin{align}\label{eq:GammaL}
      \hat{P}_{\sigma'_{sq}}(\mathbf{x})\hat{\Gamma}_{\sigma,m,n}(\mathbf{x},i)\hat{P}_0(\mathbf{x}) = \delta_{\sigma'\sigma}\delta_{sm}d_\sigma^{-1/2} \hat{\Gamma}_{\sigma,q,n}(\mathbf{x},i)\hat{P}_0(\mathbf{x}).
\end{align}
Similarly for $(\mathbf{x+i})$ at the other end of the erroneous link, 
\begin{equation}
    \hat{R}_{g^{-1}}\hat{\Gamma}_{\sigma,m,n}= \sum_{k}\Gamma^{(\sigma)}_{kn}(
g)^\ast \hat{\Gamma}_{\sigma,m,k}\hat{R}_{g^{-1}},
\end{equation}
and
\begin{align}\label{eq:GammaR}
      \hat{P}_{\sigma''_{tr}}(\mathbf{x+i})\hat{\Gamma}_{\sigma,q,n}(\mathbf{x},i)\hat{P}_0(\mathbf{x+i})
&=\delta_{\sigma''\bar{\sigma}}\delta_{tn}d_\sigma^{-1/2} \hat{\Gamma}_{\sigma,q,r}(\mathbf{x},i)\hat{P}_0(\mathbf{x+i}).
\end{align}
Combining \eq{GammaL} and \eq{GammaR} gives
\begin{align}
 \hat{P}_{\sigma''_{tr}}(\mathbf{x+i})\hat{P}_{\sigma'_{sq}}(\mathbf{x})\hat{\Gamma}_{\sigma,m,n}(\mathbf{x},i)\hat{P}_0(\mathbf{x})\hat{P}_0(\mathbf{x+i}) &= \delta_{\sigma.\sigma'}\delta_{\sigma''\bar{\sigma}}\delta_{sm}\delta_{tn}\frac{1}{d_\sigma}\hat{\Gamma}_{\sigma,q,r}(\mathbf{x},i)\hat{P}_0(\mathbf{x})\hat{P}_0(\mathbf{x+i}).
\end{align}
Using the fact $\hat{P}_0(\mathbf{x})\hat{P}_0(\mathbf{x+i})\ket{\Psi}_{\mathrm{inv}}=\ket{\Psi}_{\mathrm{inv}}$, the outcome of $\mathcal{U}_{\mathrm{dec}}$ the gauge field state $\hat{\Gamma}_{\sigma,m,n}(\mathbf{x},i)\ket{\Psi}_{\mathrm{inv}}$ is
\begin{align}
    &\sum_{\sigma'_{sq},\sigma''_{tr}}\hat{P}_{\sigma''_{tr}}(\mathbf{x+i})\hat{P}_{\sigma'_{sq}}(\mathbf{x})\hat{\Gamma}_{\sigma,m,n}(\mathbf{x},i)\ket{\Psi}_{\mathrm{inv}} \otimes \ket{\sigma'_{sq}(\mathbf{x})} \otimes \ket{\sigma''_{tr}(\mathbf{x+i})} \notag\\
    &= \sum_{q,r} \frac{1}{d_\sigma}\hat{\Gamma}_{\sigma,q,r}(\mathbf{x},i)\ket{\Psi}_{\mathrm{inv}}  \otimes \ket{\sigma'_{sq}(\mathbf{x})} \otimes \ket{\sigma''_{tr}(\mathbf{x+i})}
\end{align}
\section{Clebsch-Gordan sum in the group element basis}\label{app:CG}
  \begin{figure}
      \centering
\includegraphics[height=0.1\textwidth]{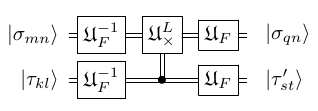}
\includegraphics[height=0.1\textwidth]{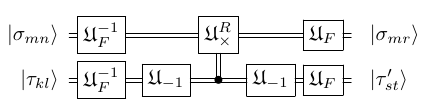}
      \caption{Quantum circuits implementing Clebsch-Gordan sums for generic finite groups. (left) $\mathfrak{U}^L_{CG}$ and (right) $\mathfrak{U}^R_{CG}$
      }
      \label{fig:U_CG}
  \end{figure}
  
The relation between Clebsch-Gordan coefficients and matrix elements of the representations \cite{Broek78CG} is :
\begin{align}\label{eq:3Gamma}
    \frac{d_{\sigma''}}{|G|} \sum_{g \in G} \Gamma^{(\sigma)}_{mn}(g) \Gamma^{(\sigma')}_{kq}(g)\Gamma^{(\sigma'')}_{lr}(g)^\ast =\sum_{\alpha}\braket{\sigma''_{r},\alpha|\sigma_n,\sigma'_{q}}\braket{\sigma_m,\sigma'_{k}|\sigma''_{l},\alpha}.
\end{align}
The Clebsch-Gordan coefficients satisfy $\braket{\sigma_n,\sigma'_{q}|\sigma''_{r},\alpha}^\ast= \braket{\bar\sigma_n,\bar\sigma'_{q}|\bar\sigma''_{r},\alpha}$. We can also choose the phases of representation to satisfy $\braket{\sigma''_{r},\alpha|\sigma_n,\sigma'_{q}}=\braket{\sigma_n,\sigma'_{q}|\sigma''_{r},\alpha}^\ast$, which will be used for later discussions.

With \eq{3Gamma}, we show that the group multiplications Fourier transformed to the representation basis is equivalent to the addition of representations according to the Clebsch-Gordan coefficients (\fig{U_CG}).
\begin{align}
   \mathfrak{U}^L_{CG}=& \frac{\sqrt{d_{\sigma}d_{\sigma'_1}d_{\tau}d_{\tau'}}}{|G|^2}\Gamma^{(\sigma')}_{qr}(g_2g_1) \Gamma^{(\tau')}_{st}(g_2) \Gamma^{(\sigma)}_{mn}(g_1)^\ast \Gamma^{(\tau)}_{kl}(g_2)^\ast \big(\ket{{\sigma'}_{qr}}\otimes \ket{{\tau'}_{st}}\bra{{\sigma}_{mn}}\otimes \bra{{\tau}_{kl}}\big) \notag\\
   =&\frac{\sqrt{d_{\sigma}d_{\sigma'_1}d_{\tau}d_{\tau'}}}{|G|^2} \Gamma^{(\sigma')}_{qu}(g_2)\Gamma^{(\sigma')}_{ur}(g_1)\Gamma^{(\tau')}_{st}(g_2) \Gamma^{(\sigma)}_{mn}(g_1)^\ast \Gamma^{(\tau)}_{kl}(g_2)^\ast \big(\ket{{\sigma'}_{qr}}\otimes \ket{{\tau'}_{st}}\bra{{\sigma}_{mn}}\otimes \bra{{\tau}_{kl}}\big) 
   \end{align}
where all the indices for group elements and group representations are summed over. Use the orthogonality \eq{ortho} to sum $g_1$, giving
\begin{align}
   \mathfrak{U}^L_{CG}&= \frac{\sqrt{d_{\tau}d_{\tau'}}}{|G|} \Gamma^{(\sigma)}_{qm}(g_2) \Gamma^{(\tau')}_{st}(g_2)\Gamma^{(\tau)}_{kl}(g_2)^\ast \big(\ket{{\sigma}_{qn}}\otimes \ket{{\tau'}_{st}}\bra{{\sigma}_{mn}}\otimes \bra{{\tau}_{kl}}\big) \notag\\
   &= \frac{\sqrt{d_{\tau}d_{\tau'}}}{|G|} \big[\Gamma^{(\bar{\sigma})}_{qm}(g_2) \Gamma^{(\tau')}_{st}(g_2)^\ast\,\Gamma^{(\tau)}_{kl}(g_2) \big]^\ast\big(\ket{{\sigma}_{qn}}\otimes \ket{{\tau'}_{st}}\bra{{\sigma}_{mn}}\otimes \bra{{\tau}_{kl}}\big),
\end{align}
Finally apply \eq{3Gamma} and convert the product of matrix elements to the Clebsch-Gordan coefficients,
\begin{align}   \mathfrak{U}^L_{CG}&= \sqrt{d_{\tau}/d_{\tau'}}\big ( \braket{{\tau'}_{t},\alpha | {\bar{\sigma}}_{m},{\tau}_{l}} \braket{{\bar{\sigma}}_{q},{\tau}_{k}| {\tau'}_{s},\alpha}\big )^\ast \big(\ket{{\sigma}_{qn}}\otimes \ket{{\tau'}_{st}} \bra{{\sigma}_{mn}}\otimes \bra{{\tau}_{kl}}\big) \notag\\
&= \sqrt{d_{\tau}/d_{\tau'}} \braket{{\bar\tau'}_{t},\alpha | \sigma_{m},\bar\tau_{l}} \braket{\sigma_{q},\bar{\tau}_{k}| {\bar{\tau}'}_{s},\alpha}\big(\ket{{\sigma}_{qn}}\otimes \ket{{\tau'}_{st}} \bra{{\sigma}_{mn}}\otimes \bra{{\tau}_{kl}}\big) 
\end{align}

Thus through $\mathfrak{U}^L_{CG}$, $\sigma$ in register 1 stays the same and $\tau$ in register 2 changes into $\tau'$. 

Thus, $\mathfrak{U}^L_{CG}$ ``adds" the left vector in register 1 to register 2 according to the Clebsch-Gordan coefficients. For register 1, the process preserves quantum numbers $\sigma, n$ but not necessarily $m$ when the representation $\sigma$ is not one-dimensional. The old flux ${{\sigma}}_{m}$ is added to ${\bar\tau}_{l}$ and the new flux ${{\sigma}}_{q}$ to ${\bar\tau}_{k}$. Similarly, $\mathfrak{U}^R_{CG}$ adds the right-vector in register 1 to register 2: 
\begin{align}
    \mathfrak{U}^R_{CG}=& \frac{\sqrt{d_{\sigma}d_{\sigma'_1}d_{\tau}d_{\tau'}}}{|G|^2}\Gamma^{(\sigma')}_{qr}(g_1g_2^{-1}) \Gamma^{(\tau')}_{st}(g_2) \Gamma^{(\sigma)}_{mn}(g_1)^\ast \Gamma^{(\tau)}_{kl}(g_2)^\ast \big(\ket{{\sigma'}_{qr}}\otimes \ket{{\tau'}_{st}}\bra{{\sigma}_{mn}}\otimes \bra{{\tau}_{kl}}\big) \notag\\
   =&\sqrt{d_{\tau}/d_{\tau'}}\braket{{\sigma}_{n},{\tau}_{l}| {\tau'}_{t},\alpha} \braket{{\tau'}_{s},\alpha | {\sigma}_{r},{\tau}_{k}} \big(\ket{{\sigma}_{mr}}\otimes \ket{{\tau'}_{st}} \bra{{\sigma}_{mn}}\otimes \bra{{\tau}_{kl}}\big) 
\end{align}

\end{widetext}
\section{Correctable errors and independence polynomials} \label{app: graph}
In this appendix, we explain our methods to compute $Q_n$ -- the number of possible ways to arrange $n $ links with correctable errors in the lattice, such that the correctability condition is still satisfied, with tools of graph theory \cite{Butera:2014nva}. The method can be applied to lattices of generic shapes. 

A graph consists of vertices and edges. An  \textit{independent subset} of vertices is a set of vertices in which no two vertices are adjacent to the same edge. The \textit{independence polynomial} of a graph is 
\begin{align}
    I(z)=\sum_{n} a_n z^n,
\end{align}
 where $a_n$ is the number of different independent subsets of $n$ vertices in the graph.
 
 For a finite lattice $\mathfrak L$, we create a corresponding graph $\mathfrak L'$ in the following way:
 
 \begin{itemize}
     \item convert each link of $\mathfrak L$ into a vertex of $\mathfrak L'$;
     \item for each pair of vertices in $\mathfrak L'$, create an edge adjacent to them, if the $\Gamma_\sigma$-type errors that occur simultaneously to the corresponding links in $\mathfrak L$ can break the correctability condition.
 \end{itemize} Thus, each way of arranging $n$ erroneous links to satisfy the correctability condition corresponds to an independent subset of vertices in the new graph. This implies that $Q_n$ is the number independent subsets of $n$ vertices in the new graph and $\Xi$ the independence polynomial.  

For each lattice considered in \fig{2dthresholds} and \fig{3dthresholds}, we construct the adjacency matrix of the corresponding graph $\mathfrak L'$ -- an $N_L\times N_L$ matrix in which the element in row $i$ and column $j$ is 0 if the $i,j$-th vertices are independent, and 1 if they are adjacent to the same edge. We then use the Python library \verb|hobj| ~\cite{Butera:2014nva} to compute the coefficients of the independence polynomials from the adjacent matrices. The connectivity for the MED condition would be higher than that for the KL condition.
\end{document}